\documentclass[letterpaper,english,reprint,aps,prl,footinbib,amssymb,floatfix,superscriptaddress,longbibliography]{revtex4-1}
\usepackage{textcomp}
\usepackage{amsmath}
\usepackage{graphicx}
\usepackage{color}

\makeatletter

\usepackage{babel}

\makeatother

\begin{document}

\title{Time resolved measurements of the switching trajectory of Pt/Co elements induced by spin-orbit torques}

\author{M. Decker$^{1}$, M. S. W{\"o}rnle$^{1}$, M. Kronseder$^{1}$, A. Meisinger$^{1}$, M. Vogel$^{1}$, H. S. K{\"o}rner$^{1}$, G. Y. Shi$^{2}$, C. Song$^{2}$,  C.H. Back}

\affiliation{Department of Physics, Regensburg University, 93053 Regensburg, Germany\linebreak $^{2}$Key Laboratory of Advanced Materials (MOE), School of Materials Science and Engineering, Tsinghua University, Beijing 100084, China}

\date{\today}

\begin{abstract}
We report the experimental observation of spin-orbit torque induced switching of perpendicularly magnetized Pt/Co elements in a time resolved stroboscopic experiment based on high resolution Kerr microscopy. Magnetization dynamics is induced by injecting sub-nanosecond current pulses into the bilayer while simultaneously applying static in-plane magnetic bias fields. Highly reproducible homogeneous switching on time scales of several tens of nanoseconds is observed. Our findings can be corroborated using micromagnetic modelling only when including a field-like torque term as well as the Dzyaloshinskii-Moriya interaction mediated by finite temperature.
\end{abstract}

\maketitle

Magnetization switching induced by spin-orbit torques (SOTs) generated by in plane (ip) current pulses in ferromagnet (FM)/heavy metal (HM) bilayers has attracted great attention in recent years~\cite{MironNature2011,LiuPRL2012,AvciAPL2012,LeeAPL2013,EmoriNatureMat2013,GarelloAPL2014,CubukcuAPL2014,BiAPL2014, LeePRB2014,LeeAPL2014,TorrejonPRB2015,YuNatureNano2014,LegrandPRA2015,ZhangAPL2015,DurrantPRB2016,FukamiNatureMat2016,LiNatureComm2016}.
A typical structure comprises a FM element with perpendicular magnetization structured on top of a HM conductor carrying the current. Technologically such a device has the advantage that the write current causing magnetization switching does not have to pass through a potential memory element itself thus avoiding its degradation~\cite{AndoJAP2014}. Studying magnetization dynamics in such elements is of interest since the exact mechanisms enabling deterministic magnetization reversal remain to be disentangled. 
SOT driven magnetization reversal in HM/FM bilayers originates from a combination of effects which manifest themselves as field and damping like torques. These torques arise from bulk and interface effects such as the bulk spin Hall effect (SHE) or the interfacial inverse spin Galvanic effect (iSGE).
Recent efforts have been dedicated to the understanding of the switching process induced by static or quasi-static currents  \cite{MironNature2011,LiuPRL2012,EmoriNatureMat2013,LeeAPL2013,LeeAPL2014,YuPRB2014,DurrantPRB2016}. 
However, the nature of the switching process itself is still under debate. 
Two possible scenarios exist: coherent rotation \cite{LeeAPL2013} or domain nucleation and propagation \cite{LiuPRL2012}. The critical current densities required for these distinct processes differ by orders of magnitude since for domain driven reversal a much smaller energy barrier needs to be overcome. 
It is believed that for devices much larger than one domain wall width, the quasi-static switching process is domain driven \cite{LiuPRL2012,EmoriNatureMat2013,LeePRB2014,DurrantPRB2016}. However, when reducing the size, it has been demonstrated recently that the switching process can be described by uniform motion \cite{ZhangAPL2015}. 
By studying switching probabilities using short current pulses of variable width \cite{AvciAPL2012,CubukcuAPL2014,GarelloAPL2014,ZhangAPL2015} reliable switching for applied pulse widths as short as 180\,ps \cite{GarelloAPL2014} has been demonstrated. In these experiments, switching dynamics is investigated indirectly by examining the final state long after the current pulse has been applied.
To understand the speed and type of the SOT induced switching process in detail, temporal and spatial resolution is required which is met in this Letter using time resolved scanning magneto-optical Kerr micoscopy (TRMOKE). 

Here we measure the trajectory of the magnetization of perpendicularly magnetized Pt/Co elements during reversal using a pump-probe approach.
We observe magnetization reversal on time scales of 10's of nanoseconds mediated by domain wall motion. By comparison with micromagnetic simulations we identify the importance of both the Dzyaloshinskii-Moriya interaction (DMI) and the field like torque for the switching process. 

\begin{figure}
	\includegraphics[width=\columnwidth]{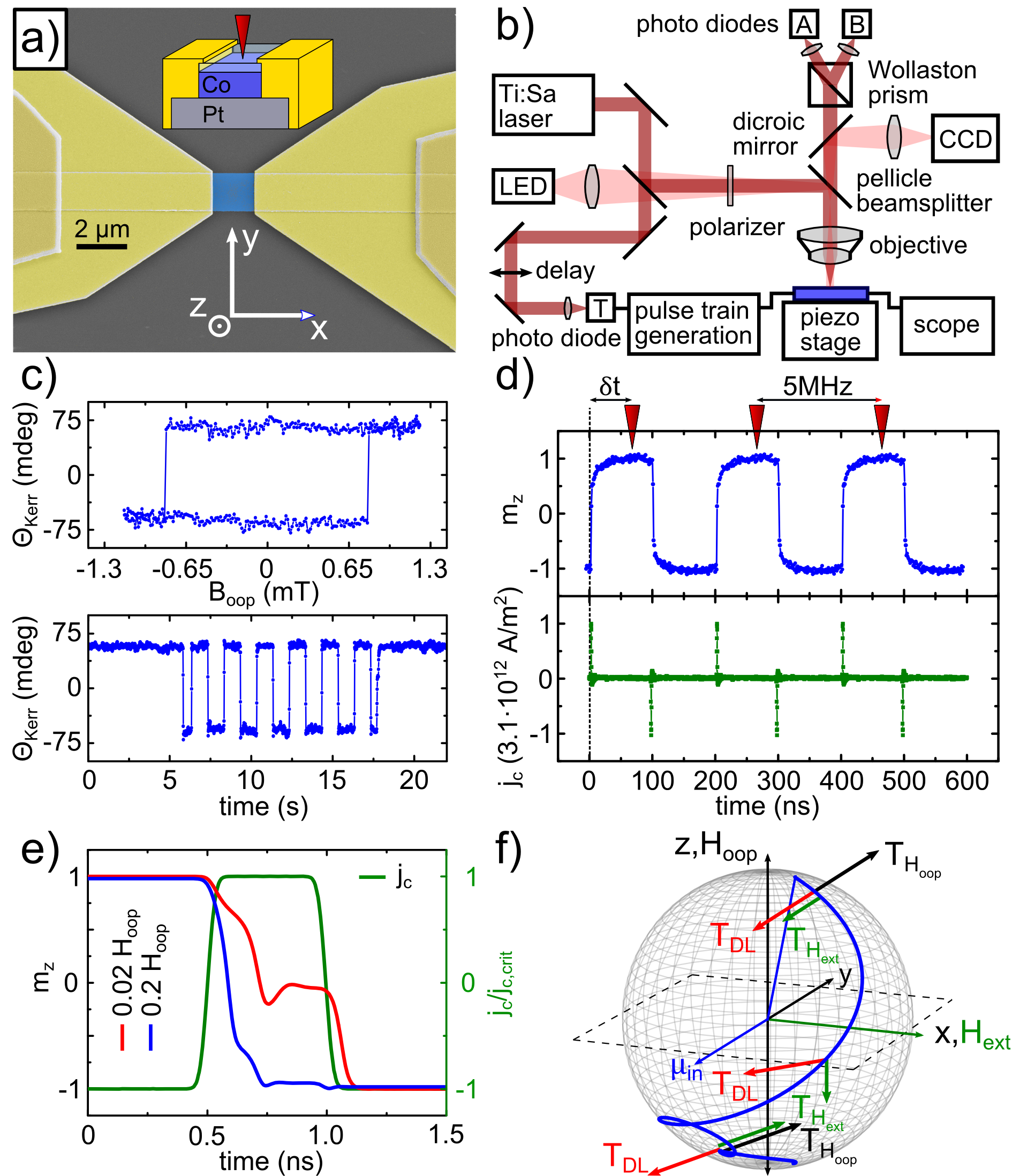}
	\caption{ a) Colored scanning electron microscopy image of the device. The 2$\times$2${\mathrm{\mu m^2}}$ Pt/Co element is marked in blue. The inset sketches the layer system. b) A simplified block diagram of the experimental setup. c) Static polar Kerr measurements recorded in the center of the element. Upper panel: hysteresis curve. Lower panel: quasi-static current induced magnetization reversal for $B_x=-20$\,mT and $j_\text{Pt}=\pm4\cdot10^{10}$\,A/m$^2$. d) Time resolved measurement of magnetization reversal for $B_x=-50$\,mT. Lower panel (green): current pulse train as transmitted through the sample. Upper panel (blue): TRMOKE signal obtained by scanning the delay $\delta t$ between current and laser pulses. The signal is recorded for one period (200\,ns) and repeated to elucidate the measurement technique. e), f): Numerical solution of the LLG for a 500\,ps current pulse (green) creating a SOT strong enough to pull the magnetization into the plane. Two cases for different external fields are shown, $H_x=+0.02\cdot H_\text{oop}$ (red) and $H_x=+0.2\cdot H_\text{oop}$ (blue). e) shows the $z$-component only while f) shows the 3D trajectory.}
	\label{fig:figure1}
\end{figure}

2\kern0.1ex\textsf{x}\kern0.1ex2\,\textmu m$^2$ Co squares on top of a 2\,\textmu m wide, 20\,\textmu m long Pt line are prepared from a stack of Ta(3nm)/\allowbreak Pt(8.5nm)/\allowbreak Co(0.5nm)/\allowbreak Al$_2$O$_3$(5nm) grown onto thermally oxidized, highly resistive silicon by molecular beam epitaxy. The square is integrated into a 50\,$\normalfont\Omega$ matched Au microstrip to ensure good transmission of short current pulses, see fig.~\ref{fig:figure1} a).
To study the dependence of size and shape, samples with 750\,nm diameter disks have been prepared on top of the 2\,\textmu m wide Pt line. The results obtained for both geometries do not differ significantly \cite{Supplement}.

To calibrate the TRMOKE experiments we record polar hysteresis loops in the center of a 2\kern0.1ex\textsf{x}\kern0.1ex2\,\textmu m$^2$ square sample in a non time resolved measurement but using the same laser system. Fig.~\ref{fig:figure1} c) shows two measurements performed statically. The upper panel shows a typical hysteresis curve.
The lower panel shows quasi-static current induced switching in an applied ip field of $B_x=-20$\,mT. The input current is a square signal of 0.5\,Hz with a current density of $\pm4\cdot10^{10}$\,A/m$^2$. Both measurements give access to the absolute Kerr signal of about 150 mdeg resulting from complete magnetization reversal enabling calibration of the signals acquired in time resolved measurements.

Magnetization dynamics is measured using a pump probe TRMOKE experiment as sketched in Fig.~\ref{fig:figure1} b) and described in detail in \cite{Supplement}. In this stroboscopic experiment the current pulse train of positive/negative pulses is generated using two pulse generators which are triggered by the laser pulses and by combining their outputs.
The resulting pulse train is then amplified by a broad band amplifier.
Using a combination of electrical and optical delay lines we adjust the delay between probing pulse and current pulse at the position of the sample with a timing jitter of $\sim$50\,ps over a full period of 200\,ns as shown in Fig.~\ref{fig:figure1} d). 
Here, 1\,ns wide pulses with a peak current density of $j_\text{max}=3.1\cdot10^{12}$\,A/m$^2$ are used to switch the magnetization in a static magnetic field of $B_x=-50$\,mT. Starting from the "down state", the positive pulse toggles the magnetization to the "up state" and the negative pulse back down, as expected for the layer sequence and a positive spin Hall angle in Pt. 

The switching process of a perpendicularly magnetized ferromagnet can be understood when solving the Landau-Lifshitz-Gilbert equation (LLG) \cite{LegrandPRA2015,LeeAPL2013,LiuPRL2012,SlonczewskiJMMM1996}
{
\begin{equation}
\begin{aligned}
\frac{\partial {\bf m}}{\partial t}
& =
-\gamma{\bf m} \times \mu_0{\boldsymbol{H}}_\text{eff}
+\alpha{\bf m} \times \frac{\partial {\bf m}}{\partial t}
\\&
+\gamma \tau_\text{DL} {\bf m}\times\left({\bf m} \times {\bf y}\right)
-\gamma \tau_\text{FL} {\bf m}\times{\bf y}.
\end{aligned}
\label{eq:LLG}
\end{equation}
}
Here $\bf{m}$ denotes the unit vector of the magnetization, $\gamma=180$\,rad/(Ts) is the gyromagnetic ratio, $\alpha$ is the Gilbert damping parameter. ${\boldsymbol{H}}_\text{eff}$ is the effective field given by the vector sum of externally applied magnetic field ${\boldsymbol{H}}_\text{ext}$ and the effective out-of-plane (oop) field ${\boldsymbol{H}}_\text{oop}=\left(\frac{2K^\perp}{\mu_0M_\text{s}}-M_\text{s}\right)m_z$ with the saturation magnetization $M_\text{s}$ and the uniaxial perpendicular anisotropy constant $K^\perp$. 
The effect of the SOTs is taken into account via the last two terms which are called damping- (${\boldsymbol{T}}_\text{DL}$) and field- (${\boldsymbol{T}}_\text{FL}$) like  SOTs, respectively. 
In the picture of the SHE, the strength of the damping-like torque can be expressed as $\tau_\text{DL}=\frac{\hbar}{2|e|}\frac{\Theta_\text{eff}j_\text{c}}{M_\text{s}d_\text{F}}$ with the effective spin Hall angle $\Theta_\text{eff}$ \cite{LegrandPRA2015,LiuPRL2012}, the ferromagnetic layer thickness $d_F$ and the injected current density $j_\text{c}||\boldsymbol{x}$. For $j_\text{c}>0$ the injected moment $\boldsymbol{ \mu}_\text{in}$ points in $-{\bf y}$ direction for our geometry \cite{HirschPRL1999}.
In the case of a full micromagnetic simulation, ${\boldsymbol{H}}_\text{eff}$ also includes the exchange interaction ($A=10^{-{11}}$J/m), and an additional DMI term which is known to be present in asymmetrically sandwiched Pt/Co/Oxide layers and is essential for reproducible domain formation~\cite{MikuszeitPRB2015}.
For deterministic bipolar switching the combination of a damping-like SOT and an additional ip field perpendicular to ${\bf y}$ is crucial~\cite{GarelloAPL2014,LiuPRL2012}. The ip field breaks the symmetry of the magnetic response to $\boldsymbol{T}_\text{DL}$ since it exerts a torque to ${\bf m}$ that adds to $\boldsymbol{T}_\text{DL}$ in one half sphere and counteracts it in the other half sphere.

In the simplest case, i.e. for coherent rotation and neglecting $\boldsymbol{T}_\text{FL}$, the switching current can be determined analytically from 
$\tau_\text{DL,crit}
=
\frac{\mu_0}{2}\left(H_\text{oop}-\sqrt{2}H_\text{ext}\right)
$~\cite{LeeAPL2013,GarelloAPL2014}.
In this case the dynamics can be calculated by numerically solving the LLG for a given set of parameters. Examples are shown in Fig.~\ref{fig:figure1} e) and f) for a 500\,ps long pulse and two different ip fields
$H_x=0.02\cdot H_\text{oop}$ (red) and $H_x=0.2\cdot H_\text{oop}$ (blue). Fig.~\ref{fig:figure1} e) shows that the magnetization switches within the first 100\,ps (250\,ps) for the larger (lower) field value. The damping-like torque is chosen such that $\tau_\text{DL,max}=\frac{\mu_0H_\text{oop}}{2}$
and $\alpha$ is set to $0.5$ \cite{MikuszeitPRB2015,MizukamiAPL2010}.
In Fig.~\ref{fig:figure1} f) the 3D trajectory of ${\bf m}(t)$ for the larger field value is shown whereas in Fig.~\ref{fig:figure1} e) the time evolution of $m_z$ is plotted.
In Fig.\ref{fig:figure1}f) the torques resulting from the fields and the SOT are shown to elucidate the role of the ip field. Note that the switching time itself strongly depends on the strength and rise time of the pulse, as worked out in \cite{LegrandPRA2015}.

\begin{figure}[h!bt]
	\includegraphics[width=1\columnwidth]{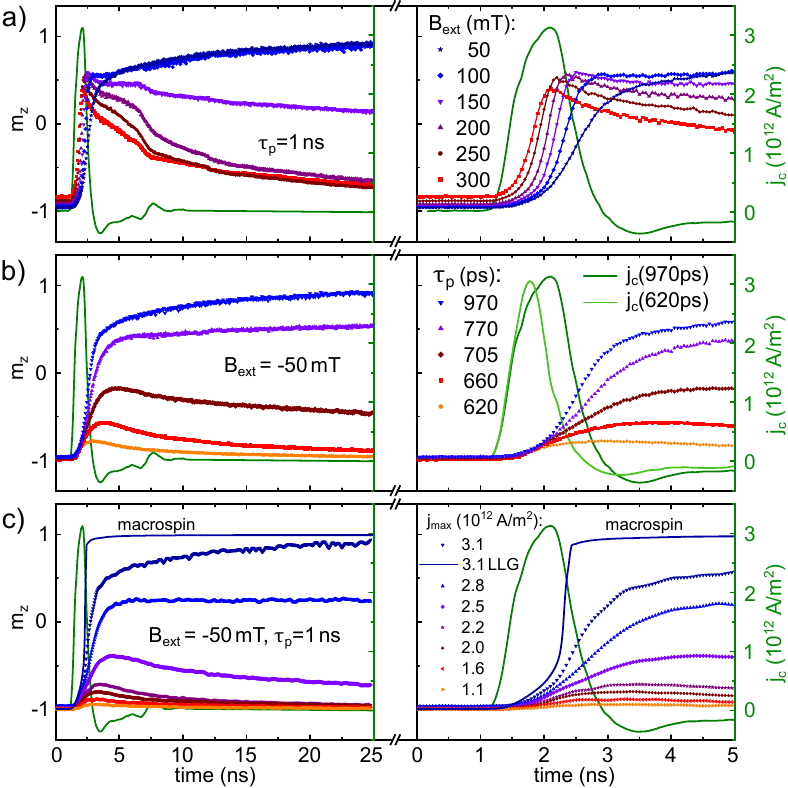}
	\caption{Transient normalized magnetization $m_z(t)$  (symbols, left axis) and  current density $j_\text{c}(t)$ (green solid lines, right axis) as function of delay time. The right graphs show the first 5\,ns of the respective left graph in detail.
	a) magnetization reversal for $\tau_\text{p}=1$\,ns and $B_\text{ext}$ ranging from $-50$\,mT to $-300$\,mT at fixed $j_\text{max}=3.1\cdot10^{12}$\,A/m$^2$. For large fields the magnetization relaxes back into the initial equilibrium position after the switching event.
	b) Measurements for fixed $B_\text{ext}=-50$\,mT and different pulse duration $\tau_\text{p}$ ranging from 0.6\,ns to 1\,ns. 
	c) Current density dependence for $\tau_\text{p}=1$\,ns and $B_\text{ext}=-50$\,mT. 
	Numerical solutions of the LLG in comparison to the experiment are shown as solid lines.
	}
	\label{fig:figure2}
\end{figure}

Simplified macrospin calculations are illustrative, but fail to explain the switching process quantitatively and even qualitatively as we will detail below. This can have multiple origins: 
i) a nonzero field-like torque $\boldsymbol{T}_\text{FL}$ reduces
the critical current; strength and sign of this torque strongly influence the dynamics~\cite{LegrandPRA2015,ZhangAPL2015}. 
ii)  it is known that switching is thermally assisted.
The influence of thermal fluctuations leads to a strong dependence of the critical current on pulse length. Note that thermal activation can be included in a macrospin calculation \cite{BedauAPL2010,LiuJMMM2014,LeeAPL2014,LedermanPRL1994}. 
iii) for magnetic elements with dimensions large enough to accomodate a domain wall, the reversal process can be driven by domain nucleation and propagation further reducing the energy barrier.
For a Pt/Co bilayer this has first been described in \cite{LiuPRL2012} where a model is used that implicitly includes thermal activation and domain driven switching for static currents. Subsequently, experiments using pulsed currents of variable pulse width $\tau_\text{p}$ have been reported discussing the effects of thermal activation \cite{AvciAPL2012,LeeAPL2014,BiAPL2014,GarelloAPL2014,ZhangAPL2015} and device size \cite{ZhangAPL2015} on the switching mechanism.

To gain insight into the mechanisms at work, we study the reversal process as a function of time. In Fig.~\ref{fig:figure2} the Kerr signal recorded in the middle of the element is displayed. The left graphs show the full recorded time span of 25\,ns; the right ones are zoomed in to the first 5\,ns only. 
Fig.~\ref{fig:figure2} a) shows the time evolution of $m_z$ for different applied fields $H_\text{ext}<H_\text{oop}$. The fixed current pulse shown in green reaches a peak current density $j_\text{max}=3.1\cdot10^{12}$\,A/m$^2$ with $\tau_\text{p}=1$\,ns. The switching process is slowest for the lowest field and speeds up as the field is increased, as expected from theory. For field values $H_\text{ext}\gg H_\text{crit}$ the magnetization switches back to the initial equilibrium position $m_z=-1$  subsequent to the pulse.
In Fig.~\ref{fig:figure2} b) the pulse width is varied between 600\,ps and 1\,ns for a fixed $B_\text{ext}=-50$\,mT and almost constant $j_\text{max}$.
For $\tau_\text{p}<1$\,ns the critical current density cannot be reached, leading to a fast decrease of the switching probability. This is seen as a reduced ``up level'' e.g. for $\tau_\text{p}=770$\,ps where the magnetization switches with about $80\%$ probability and the signal recorded is $0.8m_z-0.2m_z=0.6m_z$ due to statistical averaging of the pump probe technique. For even shorter pulses $\tau_\text{p}\leq700$\,ps, we observe only small canting of $\boldsymbol{m}$ from the equilibrium state and subsequent relaxation.
Similar results are found when reducing the peak amplitude of the current pulse at fixed $B_\text{ext}=$-50\,mT and $\tau_\text{p}=1$\,ns, see Fig.~\ref{fig:figure2} c). Additional measurements at high fields are shown in \cite{Supplement}.

\begin{figure}
	\includegraphics[width=1\columnwidth]{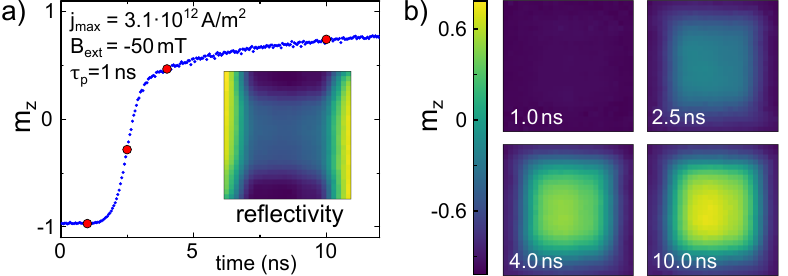}
	\caption{a)  $m_z(t)$ trace recorded in the middle of the sample, the inset shows the reflectivity (topography). b) Images of the switching process taken at distinct times marked red in a). 
	}
	\label{fig:figure3}
\end{figure}

Making use of the spatial resolution of TRMOKE, images of the switching process are taken to investigate possible spatial inhomogeneities during reversal. An example for $\tau_\text{p}=1$\,ns, $B_\text{ext}=-50$\,mT and $j_\text{max}=3.1\cdot10^{12}$\,A/m$^2$ is shown in Fig.~\ref{fig:figure3}. Fig.~\ref{fig:figure3}a) shows the switching curve in blue.
The inset shows the reflected intensity where the current contacts (yellow) can be seen on the left and right sides. The images shown in Fig.~\ref{fig:figure3}b) are snapshots of the reversal at times marked with red dots in Fig.~\ref{fig:figure3}a). The switching process is homogeneous without the appearance of propagating domains. The same result is found at different $j_\text{max}$ and external fields. In~\cite{MikuszeitPRB2015}, switching of a 100\,nm diameter disk has been simulated and it is predicted that switching occurs via deterministic domain nucleation at one side of the disk and subsequent domain wall propagation across the disk under the influence of the current. It is also shown that the nucleation point is determined by the shape of the magnetic element in combination with the presence of DMI.
Such a scenario must be visible using our experimental technique if the process is deterministic. 
To ensure that our results are not caused by the shape of the sample a disk shaped sample with a diameter of $750$\,nm was measured. The results do not differ significantly from the presented data~ \cite{Supplement}.

To obtain a quantitative understanding of the experimental results, eq.(\ref{eq:LLG}) is solved numerically using the pre-determined $M_\text{s}$ and $K^\perp$, neglecting $\boldsymbol{T}_\text{FL}$ and thermal effects, since $\tau_p\leq1$\,ns is below the thermally activated regime \cite{GarelloAPL2014,BedauAPL2010,LiuJMMM2014,LeeAPL2014}. For $\tau_\text{p}=1$\,ns and
$j_\text{c}=3.1\cdot10^{12}$\,A/m$^2$, at $B_\text{ext}=-50$\,mT, switching is achieved for a $\tau_\text{DL}$ value leading to $\Theta_\text{eff}=0.13$, which is in good agreement with published experimental results in similar structures \cite{GarelloAPL2014,ZhangNatPhys2015}, i.e. the switching current itself can in principle be explained by a simple LLG calculation.  However, the timescale of the switching process differs from the measured data as shown in Fig.~\ref{fig:figure2} c). The solid blue lines show the numerical solution corresponding to the measured data for $j_\text{max}=3.1\cdot10^{12}$\,A/m$^2$, in dark blue symbols. In the measurement the full level $m_z = 1$ is reached after $t\approx30$\,ns, in contrast to the much faster switching in the calculation. We emphasize that within the macrospin model the switching process must occur within the pulse duration due to the combination of a large value for $\alpha$ and relatively long rise/fall times of the pulse.
This statement still holds if a nonzero $\boldsymbol{T}_\text{FL}$ or thermal activation is added to the LLG.
However, 1\,ns pulses in the $10^{12}$\,A/m$^2$ regime are likely to heat up the element by some 10-100\,K which may significantly
influence the magnetic properties such as $M_\text{s}$.
To estimate the impact of a temperature increase, measurements of $M_\text{s}(T)$ 
were performed \cite{Supplement}. A Curie temperature of $T_\text{C}$ $\approx 400$\,K was found which agrees with previous reports~\cite{KoyamaAPL2015}.
To examine if Joule heating of the sample could explain the long signal recovery times, finite elements COMSOL\textregistered\,simulations
were performed to investigate how rapidly the element cools down after the pulse has passed. We find that the temperature recovers its initial value in $<$ 7\,ns after the pulse ends. Thus the magnetic signal should be recovered after this time. It should be noted that the peak temperature of the sample remains well below $T_\text{C}$ since demagnetization of the Co layer, which would manifest itself as a plateau in the Kerr data, is not observed. We conclude that a macrospin approach, even when considering Joule heating, cannot adequately explain our results. 

\begin{figure}
	\includegraphics[width=1\columnwidth]{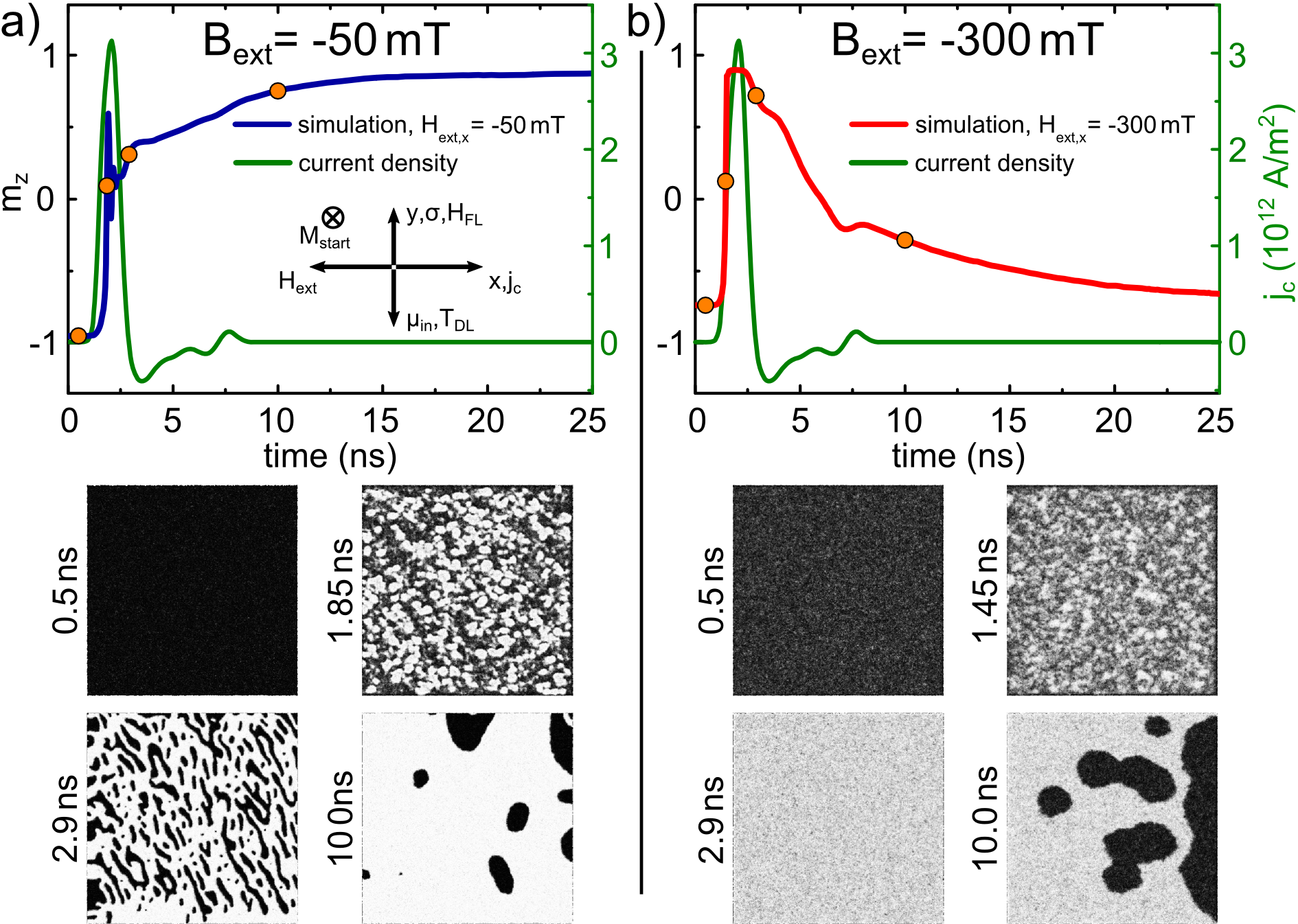}
	\caption{Micromagnetic simulations of the switching process for 2\,\textmu m wide squares using ip field values of a) -50 and b) -300\,mT to model the data shown in Fig.~\ref{fig:figure2} a). Upper panels: Time evolution of $m_\text{z}$. Orange dots denote the times for which snapshots are shown in the lower part of the figure.}
	\label{fig:figure4}
\end{figure}

To further investigate our finding that the reversal process appears homogeneous but does not follow a macrospin model, micromagnetic simulations are performed using mumax$^3$ \cite{VansteenkisteAIPAdvances2014}. The material parameters used are the same as for the LLG calculations including DMI as well. The strength of the DMI has been measured frequently in Pt/Co/Oxide systems; DMI constants range from $D\cdot d_\text{Co}\sim0.3$\,pJ/m  \cite{LeePRB2014,BenitezNatureComm2015} to 
$\sim 2$\,pJ/m\cite{KimAPL2015,PaiPRB2016,BelmeguenaiPRB2015}.
In our simulations, a DMI constant of $D\cdot d_\text{Co}=0.5$\,pJ/m is assumed.
Larger values lead to multi-domain states in zero field due to the relatively low $H_\text{oop}$ of our samples.  
In Fig. \ref{fig:figure4}a) and b) simulation results for two fields, -50 and -300\,mT, are shown.
The two upper graphs show the averaged magnetization as a function of time. The lower panels show snapshots at distinct times marked by dots in the upper traces. In the simulation the magnetic field is tilted oop by 1 degree from the $x$-axis such that at $H<0\rightarrow H_z<0$, leading to a relaxation back into the stable equilibrium for large fields, as seen in the experiment. The similarity of simulated and measured curves is striking. Especially interesting is the fact that the slow switching time can be reproduced for the $-50$\,mT case.
Most importantly we find that the experimental results can be reproduced only if DMI, a field like torque $\tau_\text{FL}$  as well as finite temperatures (100\,K) are taken into account. 

DMI in combination with finite temperatures leads to a random nucleation of small domains which, after the pulse is turned of, slowly relax into the new state by first growing and then merging. The domain pattern depends on the seed of the random field used to implement temperature. If measured with a stroboscopic technique, the statistical average over many switching events results in a homogeneous value across the sample, corroborating the results shown in Fig.~\ref{fig:figure3}. 
$\boldsymbol{T}_\text{FL}$ is necessary to reproduce the slow switching time in low fields. The sign of $\tau_\text{FL}$ is such that $\boldsymbol{T}_\text{FL}$ and $\boldsymbol{T}_\text{DL}$ try to align the magnetization in opposite directions. Without $\boldsymbol{T}_\text{FL}$ the switching process is very fast and occurs within the pulse duration. If the value of $\tau_\text{FL}$ is increased, the system is destabilized thereby continuously creating and annihilating domains during the duration of the pulse which then relax into the new state only after the pulse is turned off. The result closest to the measurement is reached for $\tau_\text{FL}/j_\text{c}=0.045$\,pTm$^2$/A and $\tau_\text{DL}/j_\text{c}=0.067$\,pTm$^2$/A. These values are close to literature values \cite{GarelloNatureNano2013}. For $|B_\text{ext}|\geq 100$\,mT the external field stabilizes the system and makes the process fast again. However, due to the low $H_\text{oop}$ and $H_\text{crit}$ in our system, the new state is not stable such that the subsequent relaxation process overlays with the switching process. 

In conclusion, we have recorded time resolved microscopic images of the SOT induced switching process of micron sized Pt/Co elements. We have shown that in order to explain our results temperature, DMI and field like torque have to be taken into account. For the case shown here switching occurs by nucleation of domains. Surprisingly and mediated by the combined action of DMI and field like torque the switching process takes much longer than the pulse duration. Note that for different parameter sets (in particular large oop anisotropies) the switching process can be described by the nucleation and propagation of a domain within the current pulse duration~\cite{MikuszeitPRB2015}.

We would like to acknowledge the DFG for funding via SFB 689 and the group of P. Gambardella for sharing their data prior to publication.


\begin{thebibliography}{35}%
\makeatletter
\providecommand \@ifxundefined [1]{%
 \@ifx{#1\undefined}
}%
\providecommand \@ifnum [1]{%
 \ifnum #1\expandafter \@firstoftwo
 \else \expandafter \@secondoftwo
 \fi
}%
\providecommand \@ifx [1]{%
 \ifx #1\expandafter \@firstoftwo
 \else \expandafter \@secondoftwo
 \fi
}%
\providecommand \natexlab [1]{#1}%
\providecommand \enquote  [1]{``#1''}%
\providecommand \bibnamefont  [1]{#1}%
\providecommand \bibfnamefont [1]{#1}%
\providecommand \citenamefont [1]{#1}%
\providecommand \href@noop [0]{\@secondoftwo}%
\providecommand \href [0]{\begingroup \@sanitize@url \@href}%
\providecommand \@href[1]{\@@startlink{#1}\@@href}%
\providecommand \@@href[1]{\endgroup#1\@@endlink}%
\providecommand \@sanitize@url [0]{\catcode `\\12\catcode `\$12\catcode
  `\&12\catcode `\#12\catcode `\^12\catcode `\_12\catcode `\%12\relax}%
\providecommand \@@startlink[1]{}%
\providecommand \@@endlink[0]{}%
\providecommand \url  [0]{\begingroup\@sanitize@url \@url }%
\providecommand \@url [1]{\endgroup\@href {#1}{\urlprefix }}%
\providecommand \urlprefix  [0]{URL }%
\providecommand \Eprint [0]{\href }%
\providecommand \doibase [0]{http://dx.doi.org/}%
\providecommand \selectlanguage [0]{\@gobble}%
\providecommand \bibinfo  [0]{\@secondoftwo}%
\providecommand \bibfield  [0]{\@secondoftwo}%
\providecommand \translation [1]{[#1]}%
\providecommand \BibitemOpen [0]{}%
\providecommand \bibitemStop [0]{}%
\providecommand \bibitemNoStop [0]{.\EOS\space}%
\providecommand \EOS [0]{\spacefactor3000\relax}%
\providecommand \BibitemShut  [1]{\csname bibitem#1\endcsname}%
\let\auto@bib@innerbib\@empty
\bibitem [{\citenamefont {Miron}\ \emph {et~al.}(2011)\citenamefont {Miron},
  \citenamefont {Garello}, \citenamefont {Gaudin}, \citenamefont {Zermatten},
  \citenamefont {Costache}, \citenamefont {Auffret}, \citenamefont {Bandiera},
  \citenamefont {Rodmacq}, \citenamefont {Schuhl},\ and\ \citenamefont
  {Gambardella}}]{MironNature2011}%
  \BibitemOpen
  \bibfield  {author} {\bibinfo {author} {\bibfnamefont {I.~M.}\ \bibnamefont
  {Miron}}, \bibinfo {author} {\bibfnamefont {K.}~\bibnamefont {Garello}},
  \bibinfo {author} {\bibfnamefont {G.}~\bibnamefont {Gaudin}}, \bibinfo
  {author} {\bibfnamefont {P.-J.}\ \bibnamefont {Zermatten}}, \bibinfo {author}
  {\bibfnamefont {M.~V.}\ \bibnamefont {Costache}}, \bibinfo {author}
  {\bibfnamefont {S.}~\bibnamefont {Auffret}}, \bibinfo {author} {\bibfnamefont
  {S.}~\bibnamefont {Bandiera}}, \bibinfo {author} {\bibfnamefont
  {B.}~\bibnamefont {Rodmacq}}, \bibinfo {author} {\bibfnamefont
  {A.}~\bibnamefont {Schuhl}}, \ and\ \bibinfo {author} {\bibfnamefont
  {P.}~\bibnamefont {Gambardella}},\ }\href@noop {} {\bibfield  {journal}
  {\bibinfo  {journal} {Nature}\ }\textbf {\bibinfo {volume} {476}},\ \bibinfo
  {pages} {189} (\bibinfo {year} {2011})}\BibitemShut {NoStop}%
\bibitem [{\citenamefont {Liu}\ \emph {et~al.}(2012)\citenamefont {Liu},
  \citenamefont {Lee}, \citenamefont {Gudmundsen}, \citenamefont {Ralph},\ and\
  \citenamefont {Buhrman}}]{LiuPRL2012}%
  \BibitemOpen
  \bibfield  {author} {\bibinfo {author} {\bibfnamefont {L.}~\bibnamefont
  {Liu}}, \bibinfo {author} {\bibfnamefont {O.~J.}\ \bibnamefont {Lee}},
  \bibinfo {author} {\bibfnamefont {T.~J.}\ \bibnamefont {Gudmundsen}},
  \bibinfo {author} {\bibfnamefont {D.~C.}\ \bibnamefont {Ralph}}, \ and\
  \bibinfo {author} {\bibfnamefont {R.~A.}\ \bibnamefont {Buhrman}},\
  }\href@noop {} {\bibfield  {journal} {\bibinfo  {journal} {Phys. Rev. Lett.}\
  }\textbf {\bibinfo {volume} {109}},\ \bibinfo {pages} {096602} (\bibinfo
  {year} {2012})}\BibitemShut {NoStop}%
\bibitem [{\citenamefont {{Onur Avci}}\ \emph {et~al.}(2012)\citenamefont
  {{Onur Avci}}, \citenamefont {Garello}, \citenamefont {{Mihai Miron}},
  \citenamefont {Gaudin}, \citenamefont {Auffret}, \citenamefont {Boulle},\
  and\ \citenamefont {Gambardella}}]{AvciAPL2012}%
  \BibitemOpen
  \bibfield  {author} {\bibinfo {author} {\bibfnamefont {C.}~\bibnamefont
  {{Onur Avci}}}, \bibinfo {author} {\bibfnamefont {K.}~\bibnamefont
  {Garello}}, \bibinfo {author} {\bibfnamefont {I.}~\bibnamefont {{Mihai
  Miron}}}, \bibinfo {author} {\bibfnamefont {G.}~\bibnamefont {Gaudin}},
  \bibinfo {author} {\bibfnamefont {S.}~\bibnamefont {Auffret}}, \bibinfo
  {author} {\bibfnamefont {O.}~\bibnamefont {Boulle}}, \ and\ \bibinfo {author}
  {\bibfnamefont {P.}~\bibnamefont {Gambardella}},\ }\href@noop {} {\bibfield
  {journal} {\bibinfo  {journal} {Appl. Phys. Lett.}\ }\textbf {\bibinfo
  {volume} {100}},\ \bibinfo {pages} {212404} (\bibinfo {year}
  {2012})}\BibitemShut {NoStop}%
\bibitem [{\citenamefont {Lee}\ \emph {et~al.}(2013)\citenamefont {Lee},
  \citenamefont {Lee}, \citenamefont {Min},\ and\ \citenamefont
  {Lee}}]{LeeAPL2013}%
  \BibitemOpen
  \bibfield  {author} {\bibinfo {author} {\bibfnamefont {K.-S. K.-J.}\
  \bibnamefont {Lee}}, \bibinfo {author} {\bibfnamefont {S.-W.}\ \bibnamefont
  {Lee}}, \bibinfo {author} {\bibfnamefont {B.-c.}\ \bibnamefont {Min}}, \ and\
  \bibinfo {author} {\bibfnamefont {K.-S. K.-J.}\ \bibnamefont {Lee}},\ }\href
  {http://scitation.aip.org/content/aip/journal/apl/102/11/10.1063/1.4798288{\%}5Cnhttp://link.aip.org/link/APPLAB/v102/i11/p112410/s1{\&}Agg=doi}
  {\bibfield  {journal} {\bibinfo  {journal} {Appl. Phys. Lett.}\ }\textbf
  {\bibinfo {volume} {102}},\ \bibinfo {pages} {112410} (\bibinfo {year}
  {2013})}\BibitemShut {NoStop}%
\bibitem [{\citenamefont {Emori}\ \emph {et~al.}(2013)\citenamefont {Emori},
  \citenamefont {Bauer}, \citenamefont {Ahn}, \citenamefont {Martinez},\ and\
  \citenamefont {Beach}}]{EmoriNatureMat2013}%
  \BibitemOpen
  \bibfield  {author} {\bibinfo {author} {\bibfnamefont {S.}~\bibnamefont
  {Emori}}, \bibinfo {author} {\bibfnamefont {U.}~\bibnamefont {Bauer}},
  \bibinfo {author} {\bibfnamefont {S.-M.}\ \bibnamefont {Ahn}}, \bibinfo
  {author} {\bibfnamefont {E.}~\bibnamefont {Martinez}}, \ and\ \bibinfo
  {author} {\bibfnamefont {G.~S.~D.}\ \bibnamefont {Beach}},\ }\href
  {http://www.ncbi.nlm.nih.gov/pubmed/23770726} {\bibfield  {journal} {\bibinfo
   {journal} {Nat. Mater.}\ }\textbf {\bibinfo {volume} {12}},\ \bibinfo
  {pages} {611} (\bibinfo {year} {2013})}\BibitemShut {NoStop}%
\bibitem [{\citenamefont {Garello}\ \emph {et~al.}(2014)\citenamefont
  {Garello}, \citenamefont {Avci}, \citenamefont {Miron}, \citenamefont
  {Baumgartner}, \citenamefont {Ghosh}, \citenamefont {Auffret}, \citenamefont
  {Boulle}, \citenamefont {Gaudin},\ and\ \citenamefont
  {Gambardella}}]{GarelloAPL2014}%
  \BibitemOpen
  \bibfield  {author} {\bibinfo {author} {\bibfnamefont {K.}~\bibnamefont
  {Garello}}, \bibinfo {author} {\bibfnamefont {C.~O.}\ \bibnamefont {Avci}},
  \bibinfo {author} {\bibfnamefont {I.~M.}\ \bibnamefont {Miron}}, \bibinfo
  {author} {\bibfnamefont {M.}~\bibnamefont {Baumgartner}}, \bibinfo {author}
  {\bibfnamefont {A.}~\bibnamefont {Ghosh}}, \bibinfo {author} {\bibfnamefont
  {S.}~\bibnamefont {Auffret}}, \bibinfo {author} {\bibfnamefont
  {O.}~\bibnamefont {Boulle}}, \bibinfo {author} {\bibfnamefont
  {G.}~\bibnamefont {Gaudin}}, \ and\ \bibinfo {author} {\bibfnamefont
  {P.}~\bibnamefont {Gambardella}},\ }\href
  {http://dx.doi.org/10.1063/1.4902443} {\bibfield  {journal} {\bibinfo
  {journal} {Appl. Phys. Lett.}\ }\textbf {\bibinfo {volume} {105}},\ \bibinfo
  {pages} {212402} (\bibinfo {year} {2014})}\BibitemShut {NoStop}%
\bibitem [{\citenamefont {Cubukcu}\ \emph {et~al.}(2014)\citenamefont
  {Cubukcu}, \citenamefont {Boulle}, \citenamefont {Drouard}, \citenamefont
  {Garello}, \citenamefont {{Onur Avci}}, \citenamefont {{Mihai Miron}},
  \citenamefont {Langer}, \citenamefont {Ocker}, \citenamefont {Gambardella},\
  and\ \citenamefont {Gaudin}}]{CubukcuAPL2014}%
  \BibitemOpen
  \bibfield  {author} {\bibinfo {author} {\bibfnamefont {M.}~\bibnamefont
  {Cubukcu}}, \bibinfo {author} {\bibfnamefont {O.}~\bibnamefont {Boulle}},
  \bibinfo {author} {\bibfnamefont {M.}~\bibnamefont {Drouard}}, \bibinfo
  {author} {\bibfnamefont {K.}~\bibnamefont {Garello}}, \bibinfo {author}
  {\bibfnamefont {C.}~\bibnamefont {{Onur Avci}}}, \bibinfo {author}
  {\bibfnamefont {I.}~\bibnamefont {{Mihai Miron}}}, \bibinfo {author}
  {\bibfnamefont {J.}~\bibnamefont {Langer}}, \bibinfo {author} {\bibfnamefont
  {B.}~\bibnamefont {Ocker}}, \bibinfo {author} {\bibfnamefont
  {P.}~\bibnamefont {Gambardella}}, \ and\ \bibinfo {author} {\bibfnamefont
  {G.}~\bibnamefont {Gaudin}},\ }\href@noop {} {\bibfield  {journal} {\bibinfo
  {journal} {Appl. Phys. Lett.}\ }\textbf {\bibinfo {volume} {104}},\ \bibinfo
  {pages} {042406} (\bibinfo {year} {2014})}\BibitemShut {NoStop}%
\bibitem [{\citenamefont {Bi}\ \emph {et~al.}(2014)\citenamefont {Bi},
  \citenamefont {Huang}, \citenamefont {Long}, \citenamefont {Liu},
  \citenamefont {Yao}, \citenamefont {Li}, \citenamefont {Huo}, \citenamefont
  {Pan},\ and\ \citenamefont {Liu}}]{BiAPL2014}%
  \BibitemOpen
  \bibfield  {author} {\bibinfo {author} {\bibfnamefont {C.}~\bibnamefont
  {Bi}}, \bibinfo {author} {\bibfnamefont {L.}~\bibnamefont {Huang}}, \bibinfo
  {author} {\bibfnamefont {S.}~\bibnamefont {Long}}, \bibinfo {author}
  {\bibfnamefont {Q.}~\bibnamefont {Liu}}, \bibinfo {author} {\bibfnamefont
  {Z.}~\bibnamefont {Yao}}, \bibinfo {author} {\bibfnamefont {L.}~\bibnamefont
  {Li}}, \bibinfo {author} {\bibfnamefont {Z.}~\bibnamefont {Huo}}, \bibinfo
  {author} {\bibfnamefont {L.}~\bibnamefont {Pan}}, \ and\ \bibinfo {author}
  {\bibfnamefont {M.}~\bibnamefont {Liu}},\ }\href
  {http://dx.doi.org/10.1063/1.4890539} {\bibfield  {journal} {\bibinfo
  {journal} {Appl. Phys. Lett.}\ }\textbf {\bibinfo {volume} {105}},\ \bibinfo
  {pages} {022407} (\bibinfo {year} {2014})}\BibitemShut {NoStop}%
\bibitem [{\citenamefont {Lee}\ \emph {et~al.}(2014{\natexlab{a}})\citenamefont
  {Lee}, \citenamefont {Liu}, \citenamefont {Pai}, \citenamefont {Li},
  \citenamefont {Tseng}, \citenamefont {Gowtham}, \citenamefont {Park},
  \citenamefont {Ralph},\ and\ \citenamefont {Buhrman}}]{LeePRB2014}%
  \BibitemOpen
  \bibfield  {author} {\bibinfo {author} {\bibfnamefont {O.~J.}\ \bibnamefont
  {Lee}}, \bibinfo {author} {\bibfnamefont {L.~Q.}\ \bibnamefont {Liu}},
  \bibinfo {author} {\bibfnamefont {C.~F.}\ \bibnamefont {Pai}}, \bibinfo
  {author} {\bibfnamefont {Y.}~\bibnamefont {Li}}, \bibinfo {author}
  {\bibfnamefont {H.~W.}\ \bibnamefont {Tseng}}, \bibinfo {author}
  {\bibfnamefont {P.~G.}\ \bibnamefont {Gowtham}}, \bibinfo {author}
  {\bibfnamefont {J.~P.}\ \bibnamefont {Park}}, \bibinfo {author}
  {\bibfnamefont {D.~C.}\ \bibnamefont {Ralph}}, \ and\ \bibinfo {author}
  {\bibfnamefont {R.~A.}\ \bibnamefont {Buhrman}},\ }\href@noop {} {\bibfield
  {journal} {\bibinfo  {journal} {Phys. Rev. B.}\ }\textbf {\bibinfo {volume}
  {89}},\ \bibinfo {pages} {024418} (\bibinfo {year}
  {2014}{\natexlab{a}})}\BibitemShut {NoStop}%
\bibitem [{\citenamefont {Lee}\ \emph {et~al.}(2014{\natexlab{b}})\citenamefont
  {Lee}, \citenamefont {Lee}, \citenamefont {Min},\ and\ \citenamefont
  {Lee}}]{LeeAPL2014}%
  \BibitemOpen
  \bibfield  {author} {\bibinfo {author} {\bibfnamefont {K.~S.}\ \bibnamefont
  {Lee}}, \bibinfo {author} {\bibfnamefont {S.~W.}\ \bibnamefont {Lee}},
  \bibinfo {author} {\bibfnamefont {B.~C.}\ \bibnamefont {Min}}, \ and\
  \bibinfo {author} {\bibfnamefont {K.~J.}\ \bibnamefont {Lee}},\ }\href@noop
  {} {\bibfield  {journal} {\bibinfo  {journal} {Appl. Phys. Lett.}\ }\textbf
  {\bibinfo {volume} {104}},\ \bibinfo {pages} {072413} (\bibinfo {year}
  {2014}{\natexlab{b}})}\BibitemShut {NoStop}%
\bibitem [{\citenamefont {Torrejon}\ \emph {et~al.}(2015)\citenamefont
  {Torrejon}, \citenamefont {Garcia-Sanchez}, \citenamefont {Taniguchi},
  \citenamefont {Sinha}, \citenamefont {Mitani}, \citenamefont {Kim},\ and\
  \citenamefont {Hayashi}}]{TorrejonPRB2015}%
  \BibitemOpen
  \bibfield  {author} {\bibinfo {author} {\bibfnamefont {J.}~\bibnamefont
  {Torrejon}}, \bibinfo {author} {\bibfnamefont {F.}~\bibnamefont
  {Garcia-Sanchez}}, \bibinfo {author} {\bibfnamefont {T.}~\bibnamefont
  {Taniguchi}}, \bibinfo {author} {\bibfnamefont {J.}~\bibnamefont {Sinha}},
  \bibinfo {author} {\bibfnamefont {S.}~\bibnamefont {Mitani}}, \bibinfo
  {author} {\bibfnamefont {J.~V.}\ \bibnamefont {Kim}}, \ and\ \bibinfo
  {author} {\bibfnamefont {M.}~\bibnamefont {Hayashi}},\ }\href@noop {}
  {\bibfield  {journal} {\bibinfo  {journal} {Phys. Rev. B.}\ }\textbf
  {\bibinfo {volume} {91}},\ \bibinfo {pages} {214434} (\bibinfo {year}
  {2015})}\BibitemShut {NoStop}%
\bibitem [{\citenamefont {Yu}\ \emph {et~al.}(2014{\natexlab{a}})\citenamefont
  {Yu}, \citenamefont {Upadhyaya}, \citenamefont {Fan}, \citenamefont {Alzate},
  \citenamefont {Jiang}, \citenamefont {Wong}, \citenamefont {Takei},
  \citenamefont {Bender}, \citenamefont {Chang}, \citenamefont {Jiang},
  \citenamefont {Lang}, \citenamefont {Tang}, \citenamefont {Wang},
  \citenamefont {Tserkovnyak}, \citenamefont {Amiri},\ and\ \citenamefont
  {Wang}}]{YuNatureNano2014}%
  \BibitemOpen
  \bibfield  {author} {\bibinfo {author} {\bibfnamefont {G.}~\bibnamefont
  {Yu}}, \bibinfo {author} {\bibfnamefont {P.}~\bibnamefont {Upadhyaya}},
  \bibinfo {author} {\bibfnamefont {Y.}~\bibnamefont {Fan}}, \bibinfo {author}
  {\bibfnamefont {J.~G.}\ \bibnamefont {Alzate}}, \bibinfo {author}
  {\bibfnamefont {W.}~\bibnamefont {Jiang}}, \bibinfo {author} {\bibfnamefont
  {K.~L.}\ \bibnamefont {Wong}}, \bibinfo {author} {\bibfnamefont
  {S.}~\bibnamefont {Takei}}, \bibinfo {author} {\bibfnamefont {S.~a.}\
  \bibnamefont {Bender}}, \bibinfo {author} {\bibfnamefont {L.-T.}\
  \bibnamefont {Chang}}, \bibinfo {author} {\bibfnamefont {Y.}~\bibnamefont
  {Jiang}}, \bibinfo {author} {\bibfnamefont {M.}~\bibnamefont {Lang}},
  \bibinfo {author} {\bibfnamefont {J.}~\bibnamefont {Tang}}, \bibinfo {author}
  {\bibfnamefont {Y.}~\bibnamefont {Wang}}, \bibinfo {author} {\bibfnamefont
  {Y.}~\bibnamefont {Tserkovnyak}}, \bibinfo {author} {\bibfnamefont {P.~K.}\
  \bibnamefont {Amiri}}, \ and\ \bibinfo {author} {\bibfnamefont {K.~L.}\
  \bibnamefont {Wang}},\ }\href {http://www.ncbi.nlm.nih.gov/pubmed/24813694}
  {\bibfield  {journal} {\bibinfo  {journal} {Nat. Nanotechnol.}\ }\textbf
  {\bibinfo {volume} {9}},\ \bibinfo {pages} {548} (\bibinfo {year}
  {2014}{\natexlab{a}})}\BibitemShut {NoStop}%
\bibitem [{\citenamefont {Legrand}\ \emph {et~al.}(2015)\citenamefont
  {Legrand}, \citenamefont {Ramaswamy}, \citenamefont {Mishra},\ and\
  \citenamefont {Yang}}]{LegrandPRA2015}%
  \BibitemOpen
  \bibfield  {author} {\bibinfo {author} {\bibfnamefont {W.}~\bibnamefont
  {Legrand}}, \bibinfo {author} {\bibfnamefont {R.}~\bibnamefont {Ramaswamy}},
  \bibinfo {author} {\bibfnamefont {R.}~\bibnamefont {Mishra}}, \ and\ \bibinfo
  {author} {\bibfnamefont {H.}~\bibnamefont {Yang}},\ }\href@noop {} {\bibfield
   {journal} {\bibinfo  {journal} {Phys. Rev. Appl}\ }\textbf {\bibinfo
  {volume} {3}},\ \bibinfo {pages} {064012} (\bibinfo {year}
  {2015})}\BibitemShut {NoStop}%
\bibitem [{\citenamefont {Zhang}\ \emph
  {et~al.}(2015{\natexlab{a}})\citenamefont {Zhang}, \citenamefont {Fukami},
  \citenamefont {Sato}, \citenamefont {Matsukura},\ and\ \citenamefont
  {Ohno}}]{ZhangAPL2015}%
  \BibitemOpen
  \bibfield  {author} {\bibinfo {author} {\bibfnamefont {C.}~\bibnamefont
  {Zhang}}, \bibinfo {author} {\bibfnamefont {S.}~\bibnamefont {Fukami}},
  \bibinfo {author} {\bibfnamefont {H.}~\bibnamefont {Sato}}, \bibinfo {author}
  {\bibfnamefont {F.}~\bibnamefont {Matsukura}}, \ and\ \bibinfo {author}
  {\bibfnamefont {H.}~\bibnamefont {Ohno}},\ }\href
  {http://dx.doi.org/10.1063/1.4926371} {\bibfield  {journal} {\bibinfo
  {journal} {Appl. Phys. Lett.}\ }\textbf {\bibinfo {volume} {107}},\ \bibinfo
  {pages} {012401} (\bibinfo {year} {2015}{\natexlab{a}})}\BibitemShut
  {NoStop}%
\bibitem [{\citenamefont {Durrant}\ \emph {et~al.}(2016)\citenamefont
  {Durrant}, \citenamefont {Hicken}, \citenamefont {Hao},\ and\ \citenamefont
  {Xiao}}]{DurrantPRB2016}%
  \BibitemOpen
  \bibfield  {author} {\bibinfo {author} {\bibfnamefont {C.~J.}\ \bibnamefont
  {Durrant}}, \bibinfo {author} {\bibfnamefont {R.~J.}\ \bibnamefont {Hicken}},
  \bibinfo {author} {\bibfnamefont {Q.}~\bibnamefont {Hao}}, \ and\ \bibinfo
  {author} {\bibfnamefont {G.}~\bibnamefont {Xiao}},\ }\href@noop {} {\bibfield
   {journal} {\bibinfo  {journal} {Phys. Rev. B.}\ }\textbf {\bibinfo {volume}
  {93}},\ \bibinfo {pages} {014414} (\bibinfo {year} {2016})}\BibitemShut
  {NoStop}%
\bibitem [{\citenamefont {Fukami}\ \emph {et~al.}(2016)\citenamefont {Fukami},
  \citenamefont {Zhang}, \citenamefont {DuttaGupta}, \citenamefont {Kurenkov},\
  and\ \citenamefont {Ohno}}]{FukamiNatureMat2016}%
  \BibitemOpen
  \bibfield  {author} {\bibinfo {author} {\bibfnamefont {S.}~\bibnamefont
  {Fukami}}, \bibinfo {author} {\bibfnamefont {C.}~\bibnamefont {Zhang}},
  \bibinfo {author} {\bibfnamefont {S.}~\bibnamefont {DuttaGupta}}, \bibinfo
  {author} {\bibfnamefont {A.}~\bibnamefont {Kurenkov}}, \ and\ \bibinfo
  {author} {\bibfnamefont {H.}~\bibnamefont {Ohno}},\ }\href
  {http://www.nature.com/doifinder/10.1038/nmat4566{\%}5Cnhttp://www.ncbi.nlm.nih.gov/pubmed/26878314}
  {\bibfield  {journal} {\bibinfo  {journal} {Nat. Mater.}\ }\textbf {\bibinfo
  {volume} {15}},\ \bibinfo {pages} {535} (\bibinfo {year} {2016})}\BibitemShut
  {NoStop}%
\bibitem [{\citenamefont {Li}\ \emph {et~al.}(2016)\citenamefont {Li},
  \citenamefont {Liu}, \citenamefont {Chang}, \citenamefont {Kalitsov},
  \citenamefont {Zhang}, \citenamefont {Csaba}, \citenamefont {Li},
  \citenamefont {Richardson}, \citenamefont {DeMann}, \citenamefont {Rimal},
  \citenamefont {Dey}, \citenamefont {Jiang}, \citenamefont {Porod},
  \citenamefont {Field}, \citenamefont {Tang}, \citenamefont {Marconi},
  \citenamefont {Hoffmann}, \citenamefont {Mryasov},\ and\ \citenamefont
  {Wu}}]{LiNatureComm2016}%
  \BibitemOpen
  \bibfield  {author} {\bibinfo {author} {\bibfnamefont {P.}~\bibnamefont
  {Li}}, \bibinfo {author} {\bibfnamefont {T.}~\bibnamefont {Liu}}, \bibinfo
  {author} {\bibfnamefont {H.}~\bibnamefont {Chang}}, \bibinfo {author}
  {\bibfnamefont {A.}~\bibnamefont {Kalitsov}}, \bibinfo {author}
  {\bibfnamefont {W.}~\bibnamefont {Zhang}}, \bibinfo {author} {\bibfnamefont
  {G.}~\bibnamefont {Csaba}}, \bibinfo {author} {\bibfnamefont
  {W.}~\bibnamefont {Li}}, \bibinfo {author} {\bibfnamefont {D.}~\bibnamefont
  {Richardson}}, \bibinfo {author} {\bibfnamefont {A.}~\bibnamefont {DeMann}},
  \bibinfo {author} {\bibfnamefont {G.}~\bibnamefont {Rimal}}, \bibinfo
  {author} {\bibfnamefont {H.}~\bibnamefont {Dey}}, \bibinfo {author}
  {\bibfnamefont {J.~S.}\ \bibnamefont {Jiang}}, \bibinfo {author}
  {\bibfnamefont {W.}~\bibnamefont {Porod}}, \bibinfo {author} {\bibfnamefont
  {S.~B.}\ \bibnamefont {Field}}, \bibinfo {author} {\bibfnamefont
  {J.}~\bibnamefont {Tang}}, \bibinfo {author} {\bibfnamefont {M.~C.}\
  \bibnamefont {Marconi}}, \bibinfo {author} {\bibfnamefont {A.}~\bibnamefont
  {Hoffmann}}, \bibinfo {author} {\bibfnamefont {O.}~\bibnamefont {Mryasov}}, \
  and\ \bibinfo {author} {\bibfnamefont {M.}~\bibnamefont {Wu}},\ }\href
  {http://www.nature.com/doifinder/10.1038/ncomms12688} {\bibfield  {journal}
  {\bibinfo  {journal} {Nat. Commun.}\ }\textbf {\bibinfo {volume} {7}},\
  \bibinfo {pages} {12688} (\bibinfo {year} {2016})}\BibitemShut {NoStop}%
\bibitem [{\citenamefont {Ando}\ \emph {et~al.}(2014)\citenamefont {Ando},
  \citenamefont {Fujita}, \citenamefont {Ito}, \citenamefont {Yuasa},
  \citenamefont {Suzuki}, \citenamefont {Nakatani}, \citenamefont {Miyazaki},\
  and\ \citenamefont {Yoda}}]{AndoJAP2014}%
  \BibitemOpen
  \bibfield  {author} {\bibinfo {author} {\bibfnamefont {K.}~\bibnamefont
  {Ando}}, \bibinfo {author} {\bibfnamefont {S.}~\bibnamefont {Fujita}},
  \bibinfo {author} {\bibfnamefont {J.}~\bibnamefont {Ito}}, \bibinfo {author}
  {\bibfnamefont {S.}~\bibnamefont {Yuasa}}, \bibinfo {author} {\bibfnamefont
  {Y.}~\bibnamefont {Suzuki}}, \bibinfo {author} {\bibfnamefont
  {Y.}~\bibnamefont {Nakatani}}, \bibinfo {author} {\bibfnamefont
  {T.}~\bibnamefont {Miyazaki}}, \ and\ \bibinfo {author} {\bibfnamefont
  {H.}~\bibnamefont {Yoda}},\ }\href {http://dx.doi.org/10.1063/1.4869828}
  {\bibfield  {journal} {\bibinfo  {journal} {J. Appl. Phys.}\ }\textbf
  {\bibinfo {volume} {115}},\ \bibinfo {pages} {172607} (\bibinfo {year}
  {2014})}\BibitemShut {NoStop}%
\bibitem [{\citenamefont {Yu}\ \emph {et~al.}(2014{\natexlab{b}})\citenamefont
  {Yu}, \citenamefont {Upadhyaya}, \citenamefont {Wong}, \citenamefont {Jiang},
  \citenamefont {Alzate}, \citenamefont {Tang}, \citenamefont {Amiri},\ and\
  \citenamefont {Wang}}]{YuPRB2014}%
  \BibitemOpen
  \bibfield  {author} {\bibinfo {author} {\bibfnamefont {G.}~\bibnamefont
  {Yu}}, \bibinfo {author} {\bibfnamefont {P.}~\bibnamefont {Upadhyaya}},
  \bibinfo {author} {\bibfnamefont {K.~L.}\ \bibnamefont {Wong}}, \bibinfo
  {author} {\bibfnamefont {W.}~\bibnamefont {Jiang}}, \bibinfo {author}
  {\bibfnamefont {J.~G.}\ \bibnamefont {Alzate}}, \bibinfo {author}
  {\bibfnamefont {J.}~\bibnamefont {Tang}}, \bibinfo {author} {\bibfnamefont
  {P.~K.}\ \bibnamefont {Amiri}}, \ and\ \bibinfo {author} {\bibfnamefont
  {K.~L.}\ \bibnamefont {Wang}},\ }\href@noop {} {\bibfield  {journal}
  {\bibinfo  {journal} {Phys. Rev. B.}\ }\textbf {\bibinfo {volume} {89}},\
  \bibinfo {pages} {104421} (\bibinfo {year} {2014}{\natexlab{b}})}\BibitemShut
  {NoStop}%
\bibitem [{Sup()}]{Supplement}%
  \BibitemOpen
  \href@noop {} {\enquote {\bibinfo {title} {{Supplementary material}},}\
  }\BibitemShut {NoStop}%
\bibitem [{\citenamefont {Slonczewski}(1996)}]{SlonczewskiJMMM1996}%
  \BibitemOpen
  \bibfield  {author} {\bibinfo {author} {\bibfnamefont {J.}~\bibnamefont
  {Slonczewski}},\ }\href@noop {} {\bibfield  {journal} {\bibinfo  {journal}
  {J. Magn. Magn. Mater.}\ }\textbf {\bibinfo {volume} {159}},\ \bibinfo
  {pages} {L1} (\bibinfo {year} {1996})}\BibitemShut {NoStop}%
\bibitem [{\citenamefont {Hirsch}(1999)}]{HirschPRL1999}%
  \BibitemOpen
  \bibfield  {author} {\bibinfo {author} {\bibfnamefont {J.~E.}\ \bibnamefont
  {Hirsch}},\ }\href {http://link.aps.org/doi/10.1103/PhysRevLett.83.1834}
  {\bibfield  {journal} {\bibinfo  {journal} {Phys. Rev. Lett.}\ }\textbf
  {\bibinfo {volume} {83}},\ \bibinfo {pages} {1834} (\bibinfo {year}
  {1999})}\BibitemShut {NoStop}%
\bibitem [{\citenamefont {Mikuszeit}\ \emph {et~al.}(2015)\citenamefont
  {Mikuszeit}, \citenamefont {Boulle}, \citenamefont {Miron}, \citenamefont
  {Garello}, \citenamefont {Gambardella}, \citenamefont {Gaudin},\ and\
  \citenamefont {Buda-Prejbeanu}}]{MikuszeitPRB2015}%
  \BibitemOpen
  \bibfield  {author} {\bibinfo {author} {\bibfnamefont {N.}~\bibnamefont
  {Mikuszeit}}, \bibinfo {author} {\bibfnamefont {O.}~\bibnamefont {Boulle}},
  \bibinfo {author} {\bibfnamefont {I.~M.}\ \bibnamefont {Miron}}, \bibinfo
  {author} {\bibfnamefont {K.}~\bibnamefont {Garello}}, \bibinfo {author}
  {\bibfnamefont {P.}~\bibnamefont {Gambardella}}, \bibinfo {author}
  {\bibfnamefont {G.}~\bibnamefont {Gaudin}}, \ and\ \bibinfo {author}
  {\bibfnamefont {L.~D.}\ \bibnamefont {Buda-Prejbeanu}},\ }\href@noop {}
  {\bibfield  {journal} {\bibinfo  {journal} {Phys. Rev. B.}\ }\textbf
  {\bibinfo {volume} {92}},\ \bibinfo {pages} {144424} (\bibinfo {year}
  {2015})}\BibitemShut {NoStop}%
\bibitem [{\citenamefont {Mizukami}\ \emph {et~al.}(2010)\citenamefont
  {Mizukami}, \citenamefont {Sajitha}, \citenamefont {Watanabe}, \citenamefont
  {Wu}, \citenamefont {Miyazaki}, \citenamefont {Naganuma}, \citenamefont
  {Oogane},\ and\ \citenamefont {Ando}}]{MizukamiAPL2010}%
  \BibitemOpen
  \bibfield  {author} {\bibinfo {author} {\bibfnamefont {S.}~\bibnamefont
  {Mizukami}}, \bibinfo {author} {\bibfnamefont {E.~P.}\ \bibnamefont
  {Sajitha}}, \bibinfo {author} {\bibfnamefont {D.}~\bibnamefont {Watanabe}},
  \bibinfo {author} {\bibfnamefont {F.}~\bibnamefont {Wu}}, \bibinfo {author}
  {\bibfnamefont {T.}~\bibnamefont {Miyazaki}}, \bibinfo {author}
  {\bibfnamefont {H.}~\bibnamefont {Naganuma}}, \bibinfo {author}
  {\bibfnamefont {M.}~\bibnamefont {Oogane}}, \ and\ \bibinfo {author}
  {\bibfnamefont {Y.}~\bibnamefont {Ando}},\ }\href@noop {} {\bibfield
  {journal} {\bibinfo  {journal} {Appl. Phys. Lett.}\ }\textbf {\bibinfo
  {volume} {96}},\ \bibinfo {pages} {152502} (\bibinfo {year}
  {2010})}\BibitemShut {NoStop}%
\bibitem [{\citenamefont {Bedau}\ \emph {et~al.}(2010)\citenamefont {Bedau},
  \citenamefont {Liu}, \citenamefont {Sun}, \citenamefont {Katine},
  \citenamefont {Fullerton}, \citenamefont {Mangin},\ and\ \citenamefont
  {Kent}}]{BedauAPL2010}%
  \BibitemOpen
  \bibfield  {author} {\bibinfo {author} {\bibfnamefont {D.}~\bibnamefont
  {Bedau}}, \bibinfo {author} {\bibfnamefont {H.}~\bibnamefont {Liu}}, \bibinfo
  {author} {\bibfnamefont {J.~Z.}\ \bibnamefont {Sun}}, \bibinfo {author}
  {\bibfnamefont {J.~A.}\ \bibnamefont {Katine}}, \bibinfo {author}
  {\bibfnamefont {E.~E.}\ \bibnamefont {Fullerton}}, \bibinfo {author}
  {\bibfnamefont {S.}~\bibnamefont {Mangin}}, \ and\ \bibinfo {author}
  {\bibfnamefont {A.~D.}\ \bibnamefont {Kent}},\ }\href@noop {} {\bibfield
  {journal} {\bibinfo  {journal} {Appl. Phys. Lett.}\ }\textbf {\bibinfo
  {volume} {97}},\ \bibinfo {pages} {262502} (\bibinfo {year}
  {2010})}\BibitemShut {NoStop}%
\bibitem [{\citenamefont {Liu}\ \emph {et~al.}(2014)\citenamefont {Liu},
  \citenamefont {Bedau}, \citenamefont {Sun}, \citenamefont {Mangin},
  \citenamefont {Fullerton}, \citenamefont {Katine},\ and\ \citenamefont
  {Kent}}]{LiuJMMM2014}%
  \BibitemOpen
  \bibfield  {author} {\bibinfo {author} {\bibfnamefont {H.}~\bibnamefont
  {Liu}}, \bibinfo {author} {\bibfnamefont {D.}~\bibnamefont {Bedau}}, \bibinfo
  {author} {\bibfnamefont {J.~Z.}\ \bibnamefont {Sun}}, \bibinfo {author}
  {\bibfnamefont {S.}~\bibnamefont {Mangin}}, \bibinfo {author} {\bibfnamefont
  {E.~E.}\ \bibnamefont {Fullerton}}, \bibinfo {author} {\bibfnamefont {J.~A.}\
  \bibnamefont {Katine}}, \ and\ \bibinfo {author} {\bibfnamefont {A.~D.}\
  \bibnamefont {Kent}},\ }\href {http://dx.doi.org/10.1016/j.jmmm.2014.01.061}
  {\bibfield  {journal} {\bibinfo  {journal} {J. Magn. Magn. Mater.}\ }\textbf
  {\bibinfo {volume} {358-359}},\ \bibinfo {pages} {233} (\bibinfo {year}
  {2014})}\BibitemShut {NoStop}%
\bibitem [{\citenamefont {Lederman}\ \emph {et~al.}(1994)\citenamefont
  {Lederman}, \citenamefont {Schultz},\ and\ \citenamefont
  {Ozaki}}]{LedermanPRL1994}%
  \BibitemOpen
  \bibfield  {author} {\bibinfo {author} {\bibfnamefont {M.}~\bibnamefont
  {Lederman}}, \bibinfo {author} {\bibfnamefont {S.}~\bibnamefont {Schultz}}, \
  and\ \bibinfo {author} {\bibfnamefont {M.}~\bibnamefont {Ozaki}},\
  }\href@noop {} {\bibfield  {journal} {\bibinfo  {journal} {Phys. Rev. Lett.}\
  }\textbf {\bibinfo {volume} {73}},\ \bibinfo {pages} {1986} (\bibinfo {year}
  {1994})}\BibitemShut {NoStop}%
\bibitem [{\citenamefont {Zhang}\ \emph
  {et~al.}(2015{\natexlab{b}})\citenamefont {Zhang}, \citenamefont {Han},
  \citenamefont {Jiang}, \citenamefont {Yang},\ and\ \citenamefont {{S. P.
  Parkin}}}]{ZhangNatPhys2015}%
  \BibitemOpen
  \bibfield  {author} {\bibinfo {author} {\bibfnamefont {W.}~\bibnamefont
  {Zhang}}, \bibinfo {author} {\bibfnamefont {W.}~\bibnamefont {Han}}, \bibinfo
  {author} {\bibfnamefont {X.}~\bibnamefont {Jiang}}, \bibinfo {author}
  {\bibfnamefont {S.-H.}\ \bibnamefont {Yang}}, \ and\ \bibinfo {author}
  {\bibfnamefont {S.}~\bibnamefont {{S. P. Parkin}}},\ }\href
  {http://dx.doi.org/10.1038/nphys3304{\%}5Cnhttp://www.nature.com/doifinder/10.1038/nphys3304{\%}5Cnhttp://www.nature.com/doifinder/10.1038/nphys3304}
  {\bibfield  {journal} {\bibinfo  {journal} {Nat. Phys.}\ }\textbf {\bibinfo
  {volume} {11}},\ \bibinfo {pages} {496} (\bibinfo {year}
  {2015}{\natexlab{b}})}\BibitemShut {NoStop}%
\bibitem [{\citenamefont {Koyama}\ \emph {et~al.}(2015)\citenamefont {Koyama},
  \citenamefont {Obinata}, \citenamefont {Hibino}, \citenamefont {Hirohata},
  \citenamefont {Kuerbanjiang}, \citenamefont {Lazarov},\ and\ \citenamefont
  {Chiba}}]{KoyamaAPL2015}%
  \BibitemOpen
  \bibfield  {author} {\bibinfo {author} {\bibfnamefont {T.}~\bibnamefont
  {Koyama}}, \bibinfo {author} {\bibfnamefont {A.}~\bibnamefont {Obinata}},
  \bibinfo {author} {\bibfnamefont {Y.}~\bibnamefont {Hibino}}, \bibinfo
  {author} {\bibfnamefont {A.}~\bibnamefont {Hirohata}}, \bibinfo {author}
  {\bibfnamefont {B.}~\bibnamefont {Kuerbanjiang}}, \bibinfo {author}
  {\bibfnamefont {V.~K.}\ \bibnamefont {Lazarov}}, \ and\ \bibinfo {author}
  {\bibfnamefont {D.}~\bibnamefont {Chiba}},\ }\href
  {http://dx.doi.org/10.1063/1.4916824} {\bibfield  {journal} {\bibinfo
  {journal} {Appl. Phys. Lett.}\ }\textbf {\bibinfo {volume} {106}},\ \bibinfo
  {pages} {132409} (\bibinfo {year} {2015})}\BibitemShut {NoStop}%
\bibitem [{\citenamefont {Vansteenkiste}\ \emph {et~al.}(2014)\citenamefont
  {Vansteenkiste}, \citenamefont {Leliaert}, \citenamefont {Dvornik},
  \citenamefont {Helsen}, \citenamefont {Garcia-Sanchez},\ and\ \citenamefont
  {{Van Waeyenberge}}}]{VansteenkisteAIPAdvances2014}%
  \BibitemOpen
  \bibfield  {author} {\bibinfo {author} {\bibfnamefont {A.}~\bibnamefont
  {Vansteenkiste}}, \bibinfo {author} {\bibfnamefont {J.}~\bibnamefont
  {Leliaert}}, \bibinfo {author} {\bibfnamefont {M.}~\bibnamefont {Dvornik}},
  \bibinfo {author} {\bibfnamefont {M.}~\bibnamefont {Helsen}}, \bibinfo
  {author} {\bibfnamefont {F.}~\bibnamefont {Garcia-Sanchez}}, \ and\ \bibinfo
  {author} {\bibfnamefont {B.}~\bibnamefont {{Van Waeyenberge}}},\ }\href
  {http://dx.doi.org/10.1063/1.4899186} {\bibfield  {journal} {\bibinfo
  {journal} {AIP Adv.}\ }\textbf {\bibinfo {volume} {4}},\ \bibinfo {pages}
  {107133} (\bibinfo {year} {2014})}\BibitemShut {NoStop}%
\bibitem [{\citenamefont {Benitez}\ \emph {et~al.}(2015)\citenamefont
  {Benitez}, \citenamefont {Hrabec}, \citenamefont {Mihai}, \citenamefont
  {Moore}, \citenamefont {Burnell}, \citenamefont {McGrouther}, \citenamefont
  {Marrows},\ and\ \citenamefont {McVitie}}]{BenitezNatureComm2015}%
  \BibitemOpen
  \bibfield  {author} {\bibinfo {author} {\bibfnamefont {M.~J.}\ \bibnamefont
  {Benitez}}, \bibinfo {author} {\bibfnamefont {A.}~\bibnamefont {Hrabec}},
  \bibinfo {author} {\bibfnamefont {a.~P.}\ \bibnamefont {Mihai}}, \bibinfo
  {author} {\bibfnamefont {T.~a.}\ \bibnamefont {Moore}}, \bibinfo {author}
  {\bibfnamefont {G.}~\bibnamefont {Burnell}}, \bibinfo {author} {\bibfnamefont
  {D.}~\bibnamefont {McGrouther}}, \bibinfo {author} {\bibfnamefont {C.~H.}\
  \bibnamefont {Marrows}}, \ and\ \bibinfo {author} {\bibfnamefont
  {S.}~\bibnamefont {McVitie}},\ }\href
  {http://www.ncbi.nlm.nih.gov/pubmed/26642936} {\bibfield  {journal} {\bibinfo
   {journal} {Nat. Commun.}\ }\textbf {\bibinfo {volume} {6}},\ \bibinfo
  {pages} {8957} (\bibinfo {year} {2015})}\BibitemShut {NoStop}%
\bibitem [{\citenamefont {Kim}\ \emph {et~al.}(2015)\citenamefont {Kim},
  \citenamefont {Han}, \citenamefont {Jung}, \citenamefont {Cho}, \citenamefont
  {Kim}, \citenamefont {Swagten},\ and\ \citenamefont {You}}]{KimAPL2015}%
  \BibitemOpen
  \bibfield  {author} {\bibinfo {author} {\bibfnamefont {N.~H.}\ \bibnamefont
  {Kim}}, \bibinfo {author} {\bibfnamefont {D.~S.}\ \bibnamefont {Han}},
  \bibinfo {author} {\bibfnamefont {J.}~\bibnamefont {Jung}}, \bibinfo {author}
  {\bibfnamefont {J.}~\bibnamefont {Cho}}, \bibinfo {author} {\bibfnamefont
  {J.~S.}\ \bibnamefont {Kim}}, \bibinfo {author} {\bibfnamefont {H.~J.~M.}\
  \bibnamefont {Swagten}}, \ and\ \bibinfo {author} {\bibfnamefont {C.~Y.}\
  \bibnamefont {You}},\ }\href {http://dx.doi.org/10.1063/1.4932550} {\bibfield
   {journal} {\bibinfo  {journal} {Appl. Phys. Lett.}\ }\textbf {\bibinfo
  {volume} {107}},\ \bibinfo {pages} {142408} (\bibinfo {year}
  {2015})}\BibitemShut {NoStop}%
\bibitem [{\citenamefont {Pai}\ \emph {et~al.}(2016)\citenamefont {Pai},
  \citenamefont {Mann}, \citenamefont {Tan},\ and\ \citenamefont
  {Beach}}]{PaiPRB2016}%
  \BibitemOpen
  \bibfield  {author} {\bibinfo {author} {\bibfnamefont {C.~F.}\ \bibnamefont
  {Pai}}, \bibinfo {author} {\bibfnamefont {M.}~\bibnamefont {Mann}}, \bibinfo
  {author} {\bibfnamefont {A.~J.}\ \bibnamefont {Tan}}, \ and\ \bibinfo
  {author} {\bibfnamefont {G.~S.~D.}\ \bibnamefont {Beach}},\ }\href@noop {}
  {\bibfield  {journal} {\bibinfo  {journal} {Phys. Rev. B.}\ }\textbf
  {\bibinfo {volume} {93}},\ \bibinfo {pages} {144409} (\bibinfo {year}
  {2016})}\BibitemShut {NoStop}%
\bibitem [{\citenamefont {Belmeguenai}\ \emph {et~al.}(2015)\citenamefont
  {Belmeguenai}, \citenamefont {Adam}, \citenamefont {Roussign{\'{e}}},
  \citenamefont {Eimer}, \citenamefont {Devolder}, \citenamefont {Kim},
  \citenamefont {Cherif}, \citenamefont {Stashkevich},\ and\ \citenamefont
  {Thiaville}}]{BelmeguenaiPRB2015}%
  \BibitemOpen
  \bibfield  {author} {\bibinfo {author} {\bibfnamefont {M.}~\bibnamefont
  {Belmeguenai}}, \bibinfo {author} {\bibfnamefont {J.~P.}\ \bibnamefont
  {Adam}}, \bibinfo {author} {\bibfnamefont {Y.}~\bibnamefont
  {Roussign{\'{e}}}}, \bibinfo {author} {\bibfnamefont {S.}~\bibnamefont
  {Eimer}}, \bibinfo {author} {\bibfnamefont {T.}~\bibnamefont {Devolder}},
  \bibinfo {author} {\bibfnamefont {J.~V.}\ \bibnamefont {Kim}}, \bibinfo
  {author} {\bibfnamefont {S.~M.}\ \bibnamefont {Cherif}}, \bibinfo {author}
  {\bibfnamefont {A.}~\bibnamefont {Stashkevich}}, \ and\ \bibinfo {author}
  {\bibfnamefont {A.}~\bibnamefont {Thiaville}},\ }\href@noop {} {\bibfield
  {journal} {\bibinfo  {journal} {Phys. Rev. B.}\ }\textbf {\bibinfo {volume}
  {91}},\ \bibinfo {pages} {180405(R)} (\bibinfo {year} {2015})}\BibitemShut
  {NoStop}%
\bibitem [{\citenamefont {Garello}\ \emph {et~al.}(2013)\citenamefont
  {Garello}, \citenamefont {Miron}, \citenamefont {Avci}, \citenamefont
  {Freimuth}, \citenamefont {Mokrousov}, \citenamefont {Bl{\"{u}}gel},
  \citenamefont {Auffret}, \citenamefont {Boulle}, \citenamefont {Gaudin},\
  and\ \citenamefont {Gambardella}}]{GarelloNatureNano2013}%
  \BibitemOpen
  \bibfield  {author} {\bibinfo {author} {\bibfnamefont {K.}~\bibnamefont
  {Garello}}, \bibinfo {author} {\bibfnamefont {I.~M.}\ \bibnamefont {Miron}},
  \bibinfo {author} {\bibfnamefont {C.~O.}\ \bibnamefont {Avci}}, \bibinfo
  {author} {\bibfnamefont {F.}~\bibnamefont {Freimuth}}, \bibinfo {author}
  {\bibfnamefont {Y.}~\bibnamefont {Mokrousov}}, \bibinfo {author}
  {\bibfnamefont {S.}~\bibnamefont {Bl{\"{u}}gel}}, \bibinfo {author}
  {\bibfnamefont {S.}~\bibnamefont {Auffret}}, \bibinfo {author} {\bibfnamefont
  {O.}~\bibnamefont {Boulle}}, \bibinfo {author} {\bibfnamefont
  {G.}~\bibnamefont {Gaudin}}, \ and\ \bibinfo {author} {\bibfnamefont
  {P.}~\bibnamefont {Gambardella}},\ }\href
  {http://www.ncbi.nlm.nih.gov/pubmed/23892985} {\bibfield  {journal} {\bibinfo
   {journal} {Nat. Nanotechnol.}\ }\textbf {\bibinfo {volume} {8}},\ \bibinfo
  {pages} {587} (\bibinfo {year} {2013})}\BibitemShut {NoStop}%
\end{thebibliography}
%

\clearpage

\onecolumngrid
\appendix

\begin{center}
{\large \bf Supplemental Material}
\end{center}

\subsection{Sample Design}   

Tow types of samples have been structured from the Ta(3nm)/\allowbreak Pt(8.5nm)/\allowbreak Co(0.5nm)/\allowbreak Al$_2$O$_3$(5nm) layers, namely a 2\kern0.1ex\textsf{x}\kern0.1ex2\,\textmu m$^2$ Pt/Co square and a 750\,nm diameter Co disk on top of a 2\,\textmu m$^2$ Pt line. 
As a first step, the \allowbreak Co/\allowbreak Al$_2$O$_3$ is etched down using wet chemical etching. As a second step the Pt line is structured using Ar+ ion etching. Subsequently a Ti(5nm)Au(250nm) microstrip is deposited in a lift-off step. The microstrip is matched to 50\,$\normalfont\Omega$ at 0.1-10\,GHz allowing for transmission of the high frequency components of the pulse. Combined with the low dc resistance of $\sim80\,\normalfont\Omega$ of the samples, the shape of the transmitted pulse is only slightly deformed and the undershoot is usually less than 10\% of the peak amplitude. The sheet resistance of the metal layer is 60$\normalfont\Omega$.

\begin{figure}[h!bt]
	\includegraphics[width=\linewidth]{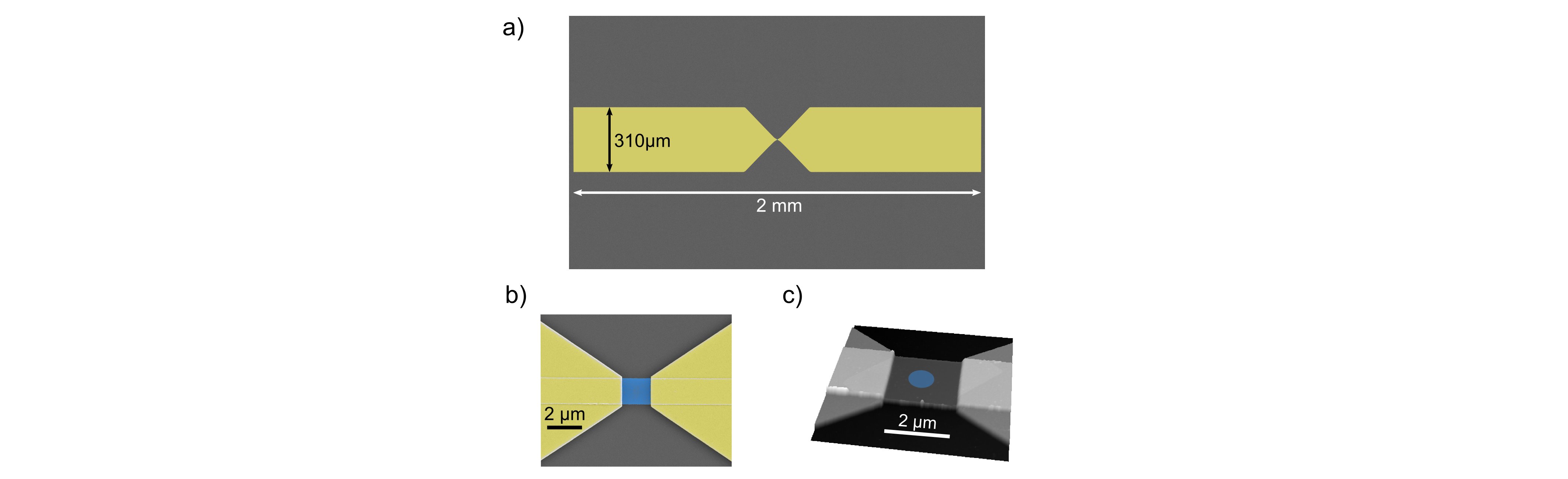}
	\caption{Sample design of the two studied devices. In a), an overview over the whole device is sketched, a 310\,\textmu m wide Au microstrip is used to guide the current pulse to the central region shown in b) and c). The two different devices studied are a 2\kern0.1ex\textsf{x}\kern0.1ex2\,\textmu m Pt/Co square, shown in b) and a Co disk having a diameter of 750\,nm on top of a 2\,\textmu m$^2$ Pt line, shown in c). The Co part is colored blue in both images. 
	}
	\label{fig:Supp_Figure_Sample}
\end{figure}

\clearpage

\subsection{Pump Probe Setup and Lock-in Technique}

\begin{figure}[h!bt]
	\includegraphics[width=1\linewidth]{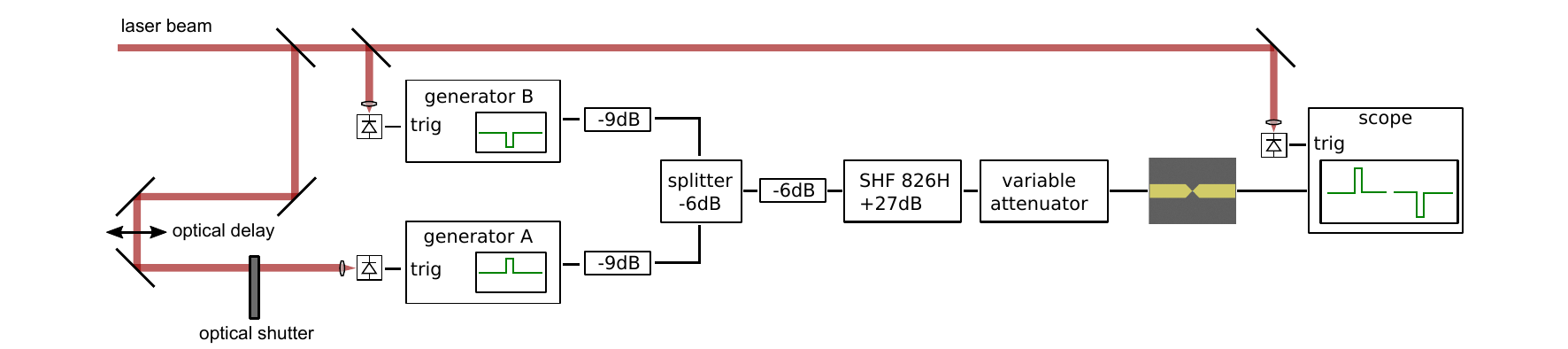}
	\caption{Detailed schematic of the pulse generation, see discussion in the text.}
	\label{fig:Supp_Figure_Pulsegen}
\end{figure}

The output beam of a modelocked Ti:Sapphire laser is split into two parts as sketched in fig.1\,b) in the main document. One part is used to probe the $z$-component of the magnetization via polar MOKE while the other triggers the current pulse train. The laser operates at 5\,MHz repetition rate and delivers $\sim$150\,fs pulses at a central wavelength of 830\,nm.
The polarized probing beam is focused onto the sample which is mounted on a $x,y,z$ piezo stage and polarization and intensity of the reflected light is measured using a Wollaston prism and a balanced photodetector, giving access to the Kerr rotation $2\Theta_\text{K}=\frac{U_\text{A}-U_\text{B}}{U_\text{A}+U_\text{B}}$. We use a 100x objective with a numerical aperture of 0.9 leading to a spatial resolution of $\sim$500nm.
A second light path (635\,nm LED) is used to record a widefield image of the sample at the same time, allowing for automated stabilization of the device under the beam. A detailed sketch of the pulse generation setup is shown in fig.\ref{fig:Supp_Figure_Pulsegen}.

The pulsetrain is generated using two HP8131A pulsegenerators and combining their outputs via a resistive powersplitter. To suppress interaction of the two generators, a 9\,dB attenuator is installed directly after each output. The pulsetrain is then amplified using a SHF 826H broadband amplifier which delivers a maximum output voltage of $\pm6$\,V. To vary the power input to the sample, a variable attenuator is used that can be set in 1\,dB steps. The pulses are then fed into the sample and the transmitted part is sampled using a 20\,GHz scope where a jitter of $~50$\,ps is usually seen as defined to be the temporal resolution in the main text. 

The time resolution is achieved in the following way:
Let A be the generator that generates the ``set pulse'', which is studied in detail, and B the generator that generates the ``reset pulse''. Each generator is triggered by a photodiode from the laser beam. An optical delay line in the trigger diode beam path of generator A allows for a delay of up to 6\,ns with a resolution of nominally 15\,fs. In addition, both generators have a programmable delay which can be set up to one full period (=200\,ns) but has a minimum stepsize of 1\,ns. Combining both delays, generator A can therefore be set to any delay with high time resolution, where the generator B can only be set in 1\,ns steps. The delays are set such that the reset pulse always comes $\approx$100\,ns after the set pulse and usually only the response to the set pulse is measured using the fine time resolution. 

The measurement over one full period shown in fig.1d) in the main document has been recorded using a 1\,ns timestep and directly reading out the Kerr signal from the photo detector. To study one switching process in detail, a Lockin technique is used to greatly enhance the signal to noise and reduce the measurement time. To modulate the signal, an optical shutter is placed in the trigger beam path of generator A, chopping the beam at 1.6\,kHz. Therefore, if the beam is blocked, only the reset pulse enters the sample and defines the equilibrium state of the magnetization. If the beam is not blocked, the signal of the magnetic response to the set pulse is recorded. 

\clearpage

\subsection{High Field Data}   

\begin{figure}[h!bt]
	\includegraphics[width=\linewidth]{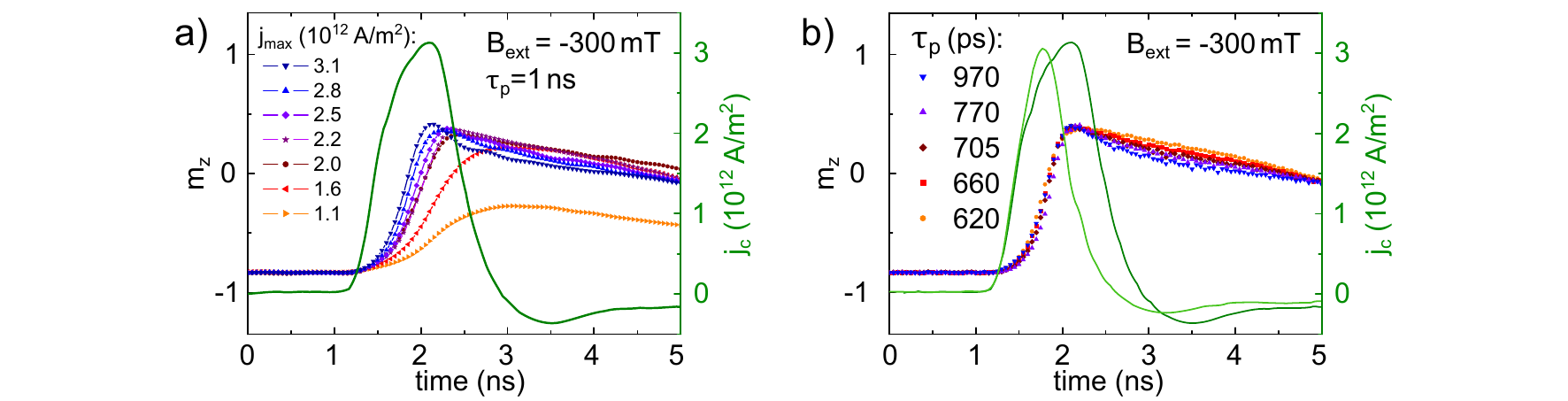}
	\caption{ 
	Transient normalized magnetization $m_z(t)$  (symbols, left axis) and  current density $j_\text{c}(t)$ (green solid lines, right axis) as function of delay time for a high applied field $B_\text{ext}=-300$\,mT.
	a) Current density dependence for $\tau_\text{p}=1$\,ns. 
	b) Measurements for different pulse duration $\tau_\text{p}$ ranging from 0.6\,ns to 1\,ns. 
	} 
	\label{fig:Supp_Figure_TR_Switch_HighFields}
\end{figure}

In addition to the data presented in Fig.~2 in the main document, Fig.~\ref{fig:Supp_Figure_TR_Switch_HighFields} shows mesurements of the power dependence in panel a) and the pulsewidth dependence in panel b) for a large applied field of $B_\text{ext}$=-300\,mT. In the left graph, showing the power dependence, one can see that, in contrast to the low field data, switching is achieved for current densities as small as 1.6$\cdot10^{12}$\,A/m$^2$. The process is faster for higher power, as expected. Unfortunately, due to the low $H_\text{oop}$ and $H_\text{crit}$ in our system, the magnetization falls back into the stable equilibrium directly after the pulse passes. However, this fast back-switching would be impossible in the macrospin case and therefore gives another direct hint to the domain driven nature of the whole process. From the pulsewidth dependence (right graph) one can see that at the large field value switching occurs for every pulse width. Here, the magnetization traces do not differ much during the switching up but after reaching the highest $m_z$-value. The curves with longer pulse width decay faster than the ones caused by a shorter pulse. This can be understood by joule heating, which is stronger for longer pulses and favors domain nucleation as well as it reduces $M_\text{s}$ which proportionally also reduces the measured Kerr signal. 

\clearpage

\subsection{Experimental Results for a Disk-shaped Element}   

\begin{figure}[h!bt]
	\includegraphics[width=0.95\linewidth]{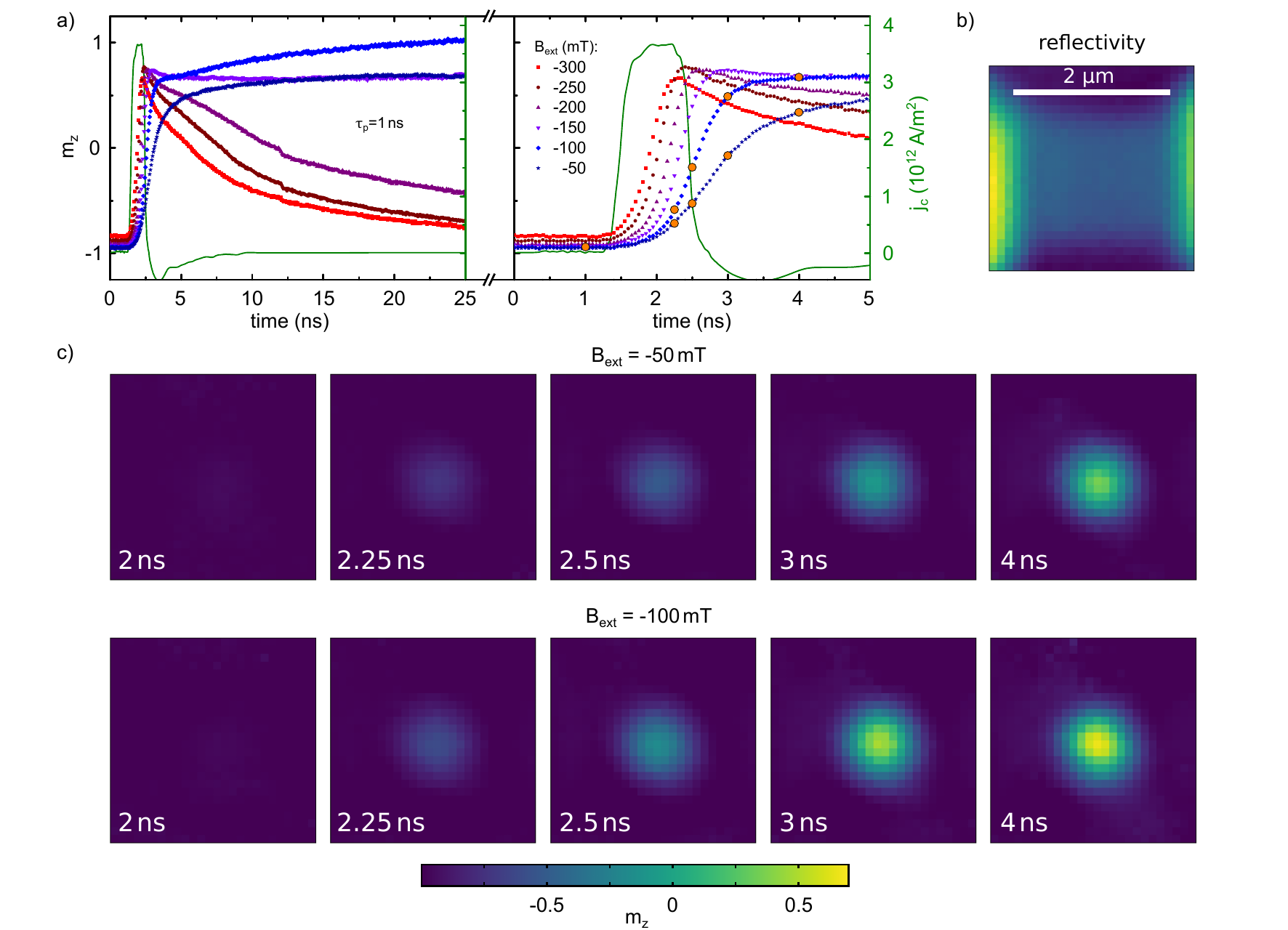}
	\caption{ 
	Transient normalized magnetization $m_z(t)$  (symbols, left axis) and  current density $j_\text{c}(t)$ (green solid lines, right axis) as function of delay time for a high applied field $B_\text{ext}=-300$\,mT.
	a) magnetization reversal for $\tau_\text{p}=1$\,ns and $B_\text{ext}$ ranging from $-50$\,mT to $-300$\,mT at fixed $j_\text{max}=3.1\cdot10^{12}$\,A/m$^2$. The right graphs show the first 5\,ns of the respective left graph in detail.  For large fields the magnetization relaxes back into the initial equilibrium position after the switching event. 
	b) topography of the device as recorded by the reflective signal.
	c) snapshots of the magnetization at distinct times denoted by orange dots in the right graph of panel a). 
	} 
	\label{fig:Supp_Figure_Disk_Sample_Res}
\end{figure}

As discussed in the main document, the shape of the magnetic element is expected to play a role in the switching process. In \cite{MikuszeitPRB2015} it is shown that for a Pt/Co disk of 100\,nm in diameter the switching happens due to deterministic domain nucleation and propagation. The main features that are critical for this behavior are the circular shape of the magnetic element and the DMI in combination with the ip applied field. To examine the effect of the shape on the switching process, a sample consisting of a 750\,nm diameter Co disk on top of a 2\,\textmu m Pt line, as shown in Fig.~\ref{fig:Supp_Figure_Sample} was studied in comparison to the 2\kern0.1ex\textsf{x}\kern0.1ex2\,\textmu m$^2$ square samples. It is found that the general behavior is the same in both cases as shown in Fig.~\ref{fig:Supp_Figure_Disk_Sample_Res}. Panel a) shows the switching curves as a function of the applied ip field in comparison to Fig.~2 of the main document. The left graphs show the full recorded time span of 25\,ns; the right ones are zoomed in to the first 5\,ns only.  Panel b) shows the measured reflective signal, where the Au contacts are at the left and right edge (yellow) and the Pt line can be seen in the center of the image. The Co disk cannot be seen in the reflective signal since it is too thin (0.5\,nm). To find the center of the disk, images are recorded at high delay times where the magnetic signal is high and subsequently the position is fixed using the Au contacts as landmarks. 
Panel c) shows snapshots of the magnetization at distinct times that are denoted by orange dots in the right graph of panel a). The upper row is recorded at $B_\text{ext}$=-50\,mT whereas the lower row is recorded at $B_\text{ext}$=-100\,mT. It can be seen that the switching process is homogeneous in both cases. We therefore conclude that the shape does not have a big influence on the switching in our case. 


\subsection{Micromagnetic Simulations}

\begin{figure}[h!bt]
	\includegraphics[width=\linewidth]{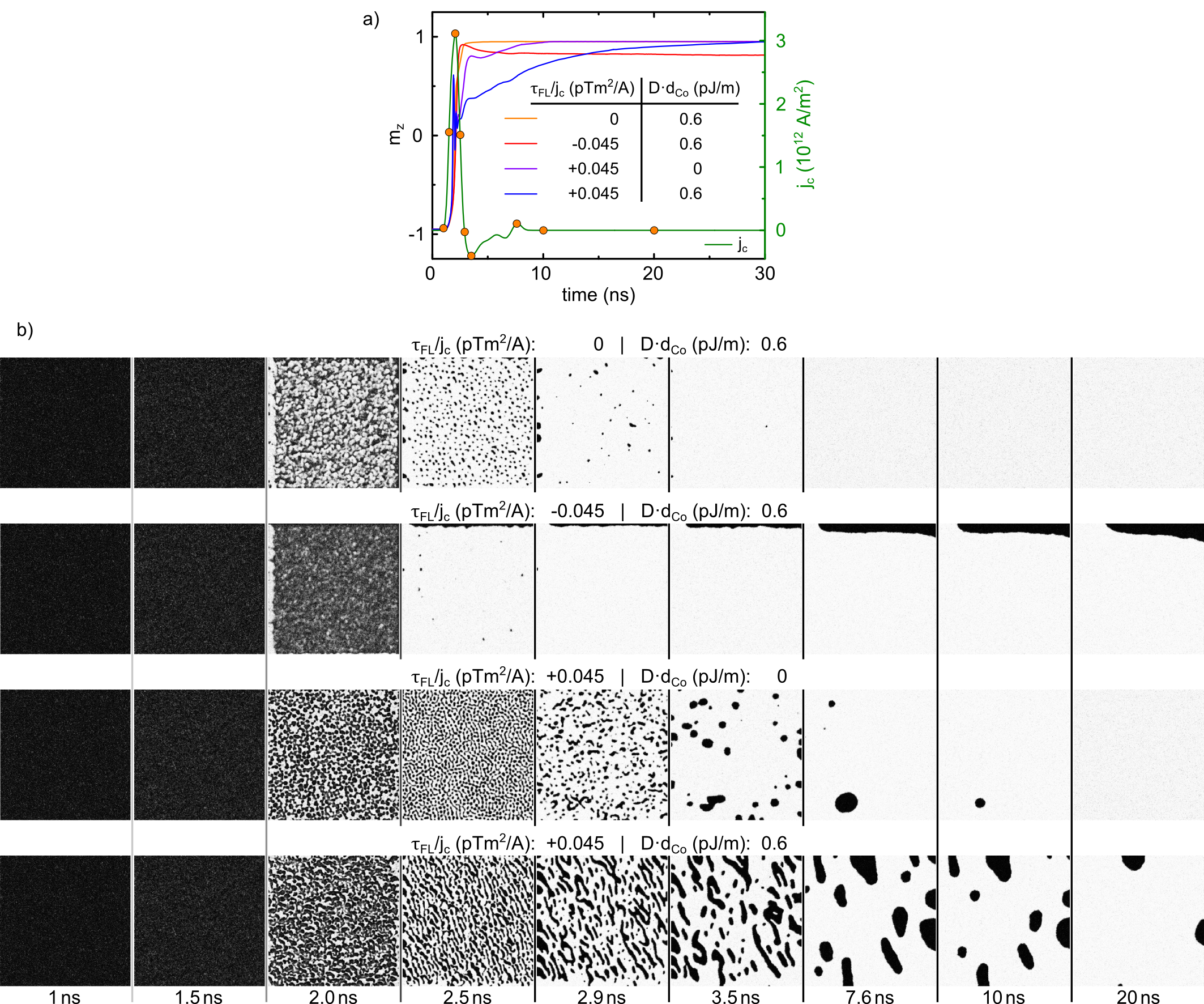}
	\caption{ 
	Simulation Data of the switching process in the 2\kern0.1ex\textsf{x}\kern0.1ex2\,\textmu m$^2$ Pt/Co square for an applied field of $B_\text{ext}=-50$\,mT and $\tau_\text{p}=1$\,ns. Shown are results for different directions of the field like torque and in dependence of DMI. Panel a) shows the averaged magnetization of the element as function of time (left axis) and the current pulse used for the simulation (right axis). Panel b) shows snapshots of $m_z$ at distinct times denoted by orange dots in panel a).   
	}
	\label{fig:Supp_Figure_Simulation}
\end{figure}

Micromagnetic simulation were carried out using the mumax$^3$ package \cite{VansteenkisteAIPAdvances2014}. All simulations of the 2\kern0.1ex\textsf{x}\kern0.1ex2\,\textmu m$^2$ Co square element ($d_\text{Co}$=0.5\,nm) were calculated on a grid of 512\kern0.1ex\textsf{x}\kern0.1ex512\kern0.1ex\textsf{x}\kern0.1ex1 cells, leading to a lateral cell size of $w_\text{cell}\approx$3.9\,nm. Material parameters used were: saturation magnetization $M_\text{s}$=1050\,kA/m, uniaxial oop anisotropy energy density $K^\perp$=971\,kJ/m$^3$, Gilbert damping constant $\alpha$=0.5 and a DMI constant of D=1\,mJ/m$^2$. 
To introduce thermal fluctuations a constant temperature of $T$=100\,K was set.
At zero Kelvin switching always happens by deterministic domain nucleation at the edge and subsequent propagation as described in \cite{MikuszeitPRB2015}. 
The value of $T$ was chosen such that $M_\text{s}$ is still mostly unaltered. 
The field an damping like torques were chosen following literature values for a similar layer structure \cite{GarelloNatureNano2013}.
The sign of the field like torque as well as the presence of DMI were varied to reveal the influence on the switching process as shown in Fig.~\ref{fig:Supp_Figure_Simulation}.  
In panel a) the average magnetization of the element is plotted as function of time (left axis) together with the current pulse used in the simulation (right axis) for different cases of sign/presence of $\tau_\text{FL}$ and DMI. 
Due to the finite temperature the switching process in all cases happens due to formation of small domains that take more or less time to relax into the new state after the pulse has passed, depending on the chosen parameter set. It can be seen that a combination of nonzero DMI and a positive $\tau_\text{FL}$ results in a very long switching times as seen in the experiment. 
Panel b) of the figure shows snapshots of $m_z$ for distinct times that are denoted by orange dots in a).


\subsection{Squid Data}   

\begin{figure}[h!bt]
	\includegraphics[width=\linewidth]{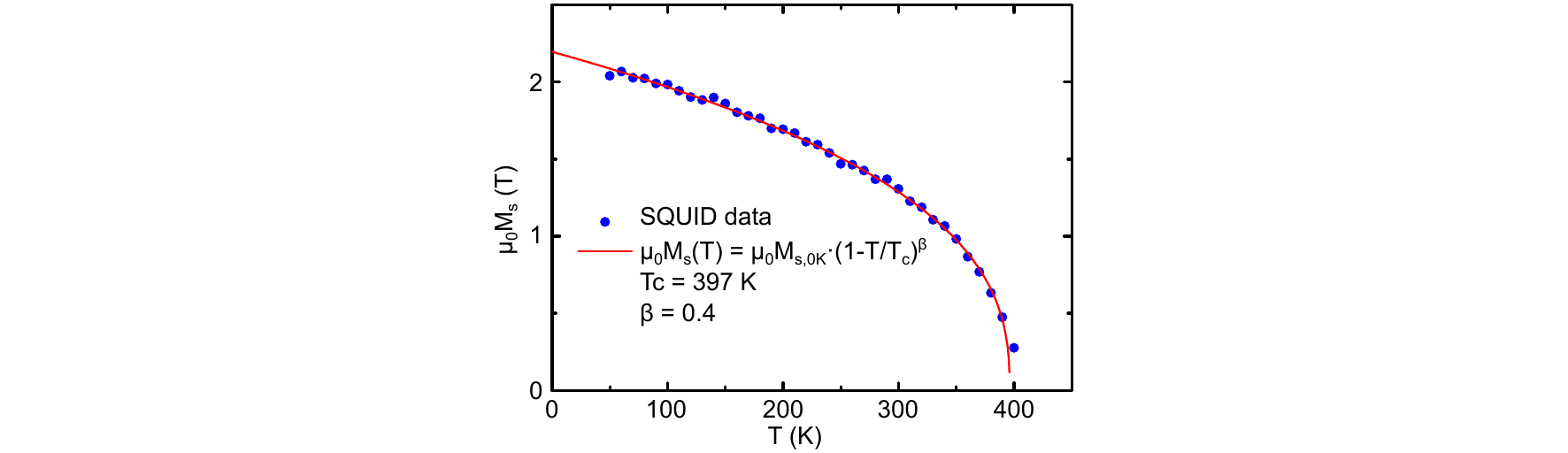}
	\caption{ 
	Temperature dependence of the saturation magnetization measured by SQUID.  
	} 
	\label{fig:Supp_Figure_SQUID}
\end{figure}

To determine the Curie temperature of our layer stack, the saturation magnetization of a full film was measured as function of temperature using a superconducting quantum interference device (SQUID) in a range of 50-400\,K. From the results shown in Fig.~\ref{fig:Supp_Figure_SQUID} a Curie temperature $T_\text{C}\approx400$\,K is determined. 


\end{document}